\documentclass[]{spie}

\usepackage{aas_macros,amsmath,amsfonts,amssymb,bm,graphicx,xcolor}
\usepackage[colorlinks=true,allcolors=blue]{hyperref}

\usepackage{mathrsfs}

\let\originalleft\left
\let\originalright\right
\renewcommand{\left}{\mathopen{}\mathclose\bgroup\originalleft}
\renewcommand{\right}{\aftergroup\egroup\originalright}

\newcommand{\ed}{\mathop{}\!\mathrm{d}}
\newcommand{\ab}[1]{\left|#1\right|}
\newcommand{\br}[1]{\left[#1\right]}
\newcommand{\cu}[1]{\left\{#1\right\}}
\newcommand{\pa}[1]{\left(#1\right)}

\definecolor{red}{RGB}{228,26,28}
\definecolor{blue}{RGB}{55,126,186}
\definecolor{green}{RGB}{77,175,74}
\definecolor{darkpurple}{RGB}{152,78,163}
\definecolor{lightpurple}{RGB}{230,210,233}

\title{The Black Hole Explorer:\\
Photon Ring Science, Detection and Shape Measurement}

\author[1]{\small Alexandru Lupsasca}
\author[2]{Alejandro C\'ardenas-Avenda\~no}
\author[3,4]{Daniel C. M. Palumbo}
\author[3,4]{Michael D. Johnson}
\author[5]{\vspace{10pt}\newline Samuel E. Gralla}
\author[6]{Daniel P. Marrone}
\author[3]{Peter Galison}
\author[3,4]{Paul Tiede}
\author[2]{Lennox Keeble}

\affil[1]{\small Department of Physics \& Astronomy, Vanderbilt University, Nashville, TN 37212, USA}
\affil[2]{Department of Physics, Princeton University, Princeton, New Jersey 08544, USA}
\affil[3]{Black Hole Initiative at Harvard University, Cambridge, MA 02138, USA}
\affil[4]{Center for Astrophysics $|$ Harvard \& Smithsonian, Cambridge, MA 02138, USA}
\affil[5]{Department of Physics, University of Arizona, Tucson, AZ 85721, USA}
\affil[6]{Steward Observatory, University of Arizona, Tucson, AZ 85721, USA}

\authorinfo{Corresponding Author: Alexandru Lupsasca (\url{alexandru.v.lupsasca@vanderbilt.edu})}

\begin{document} 
\maketitle

\begin{abstract}
% 250 words maximum
General relativity predicts that black hole images ought to display a bright, thin (and as-of-yet-unresolved) ring.
This ``photon ring'' is produced by photons that explore the strong gravity of the black hole, flowing along trajectories that experience extreme light bending within a few Schwarzschild radii of the horizon before escaping.
The shape of the photon ring is largely insensitive to the precise details of the emission from the astronomical source surrounding the black hole and therefore provides a direct probe of the Kerr geometry and its parameters.
The Black Hole Explorer (BHEX) is a proposed space-based experiment targeting the supermassive black holes M87* and Sgr\,A* with radio-interferometric observations at frequencies of 100\,GHz through 300\,GHz and from an orbital distance of $\lesssim$~30,000\,km.
This design will enable measurements of the photon rings around both M87* and Sgr\,A*, confirming the Kerr nature of these sources and delivering sharp estimates of their masses and spins.
\end{abstract}

% Include a list of keywords after the abstract (up to 8) 

\clearpage

\section{Introduction}
\label{sec:Introduction}

General relativity (GR) predicts that images of a black hole ought to display a narrow ``photon ring'' consisting of multiple mirror images of the main emission from surrounding matter.
These mirror images arise from photons that circumnavigate the black hole a different number of times on their way from their source to observer, probing the warped space-time geometry just outside the event horizon.
This orbiting light carries information about the strong gravitational field of the black hole, which is encoded in the observable shape of the photon ring.

A measurement of this predicted (but not yet observed) ring would thus provide a precise probe of strong-field gravity and will be one of the primary objectives of a NASA mission proposed to fly within the next decade: the Black Hole Explorer (BHEX).
This mission will not only resolve the photon ring, but also measure its shape: BHEX will be specially designed to isolate the interferometric signature of the photon ring (a characteristic periodic oscillation in the radio visibility), from which the angle-dependent ring diameter can be extracted and compared against the general-relativistic Kerr prediction.
After confirming the Kerr nature of a black hole source, the parameters of the black hole can be fitted to the shape of its photon ring, with the black hole mass and spin roughly corresponding to the overall size and eccentricity of its ring, respectively.

BHEX will specifically target the photon rings around the supermassive black holes M87* and Sgr\,A*.
Each source imposes its own orbital requirements on the mission, and the choice of orbit is crucial for enabling BHEX to successfully detect these photon rings and measure their shapes.
Radio observations using very-long-baseline interferometry (VLBI) with a space element at an orbital distance of $\sim$30,000\,km, at an envisioned frequency of 345\,GHz and with a 32\,GHz bandwidth, will grant access, from each space-ground baseline, to a visibility window containing about one full period of the characteristic ringing produced by the photon ring.

A series of companion papers provide additional details on BHEX, including a discussion of the motivation and larger vision for the mission \cite{Johnson2024}, an overview of its instrument system \cite{Marrone2024}, and a description of its planned operation as a hybrid ground-space observatory \cite{Issaoun2024}.
Here, we focus on the science behind the photon ring and how it informs the basic design of the mission.
We begin in Sec.~\ref{sec:RingImage} with a review of the photon ring and its GR-predicted structure in the image domain.
Then, in Sec.~\ref{sec:RingVisibility}, we describe the characteristic interferometric signature that the photon ring produces in the visibility domain and that is directly accessible to VLBI observations.
Next, in Sec.~\ref{sec:Model}, we examine a simple semi-analytic model of M87* and simulate an idealized measurement of its photon ring shape.
Finally, in Sec.~\ref{sec:GRMHD}, we present a state-of-the-art simulation of M87* based on general-relativistic magnetohydrodynamics (GRMHD) and forecast one night of observations of the source with BHEX.
These examples give an overview of the challenges involved in a photon ring shape measurement with the baseline coverage envisioned for BHEX.
We conclude in Sec.~\ref{sec:Discussion} with a discussion of outstanding problems and prospects for photon ring science with BHEX.

\section{The photon ring in the image domain}
\label{sec:RingImage}

In this section, we briefly review the origin and structure of the photon ring in black hole images.
The appearance of this ring can ultimately be traced back to the existence around any black hole (and as far as we know, only around a black hole) of a region of spacetime known as the ``photon shell'' in which light can be (unstably) trapped on spherical photon orbits.
This property was first discovered for the non-rotating (Schwarzschild) black hole by David Hilbert in 1916 \cite{Hilbert1917}, the very same year that this first black hole solution to Einstein's theory was published---Hilbert also computed the apparent size of what we now call the black hole shadow (see below).

Since light can orbit in the photon shell of the black hole, any two spatial points in the surrounding spacetime are connected by infinitely many light rays (null geodesics), each one differing by the number of orbits it executes within the shell.
As a result, an isotropic point source produces multiple mirror images, with each image arising from photons that traveled along a different ray connecting source to observer.
This lensing behavior was first analyzed in 1959 by Darwin \cite{Darwin1959} and then brought to light in 1979 by Luminet \cite{Luminet1979}, who produced the first simulated image of a black hole with a thin accretion disk: Fig.~11 in Luminet's paper clearly displays a mirror image of the disk that takes the form of a ring encircling the black hole, and which today we would dub the photon ring.

Until 2018, the photon ring received only a handful of mentions annually, but following the 2019 release by the Event Horizon Telescope (EHT) collaboration of the first horizon-scale image of the black hole in M87 \cite{EHT2019a}, interest in the photon ring exploded.
Within the past five years alone, almost 1,000 papers have been published on the photon ring and the pace is still accelerating, with $\sim1$ new paper on the topic coming out every day now.

Given the rapid growth of the field, it is no longer possible to provide an exhaustive review of the literature on the photon ring.
Here, we only cover the basic concepts most relevant to BHEX, and delegate all detailed explanations to the supplied references.
For a beginner's guide to the topic of black hole imaging with a special emphasis on the photon ring, we direct the reader to the recent pedagogical introduction by Lupsasca et al. \cite{Lupsasca2024}

The main takeaways from this section are:
\begin{itemize}
    \item General relativity predicts that embedded within a black hole image, there lies a thin photon ring composed of a sequence of self-similar subrings \cite{Luminet1979}.
    \item Each subring is a lensed image of the main emission, indexed by the number $n$ of photon half-orbits executed around the black hole \cite{Gralla2019,JohnsonLupsasca2020}.
    \item The lensing behavior of a Kerr black hole is completely characterized by three critical exponents $\gamma$, $\delta$, and $\tau$ that respectively control the demagnification, rotation and time delay of successive images \cite{JohnsonLupsasca2020,GrallaLupsasca2020a}.
    \item The photon subrings converge to (but are distinct from) a specific shape predicted by the Kerr geometry (known as the ``critical curve'') and therefore, a photon ring shape measurement yields information about the black hole mass, spin, and inclination.
\end{itemize}
In the next section, we will describe the characteristic interferometric signature that is produced by the photon ring and its subring structure and whose detection will be the target of space-VLBI observations with BHEX.

\subsection{The photon shell of spherical bound orbits}

\begin{figure}[t]
    \centering
    \includegraphics[width=.32\textwidth]{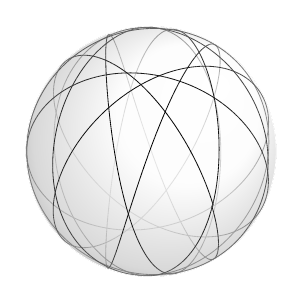}
    \includegraphics[width=.32\textwidth]{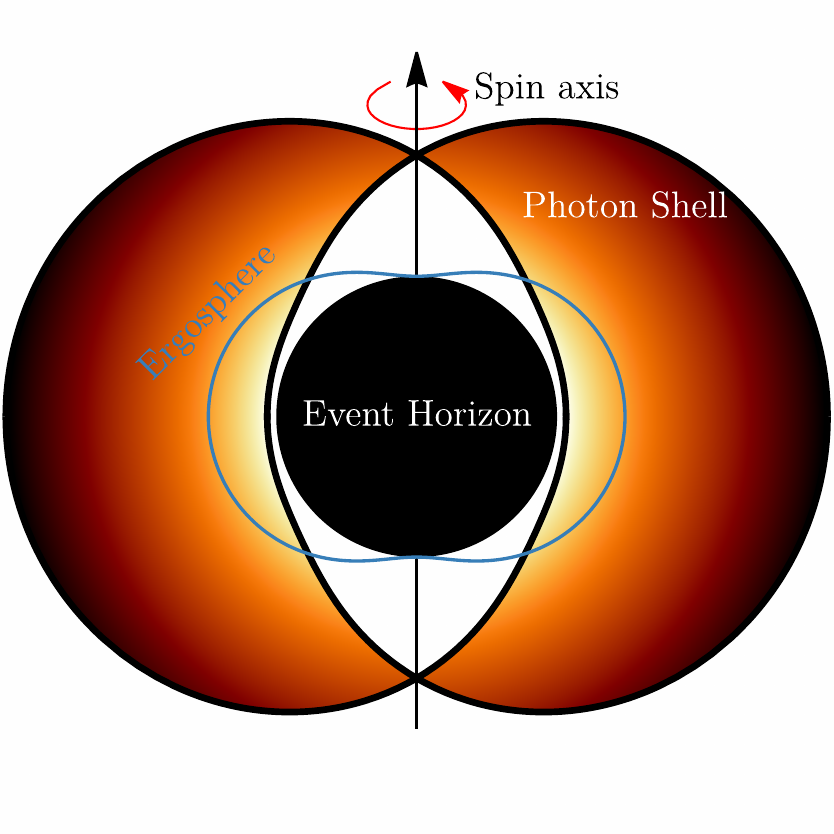}
    \includegraphics[width=.32\textwidth]{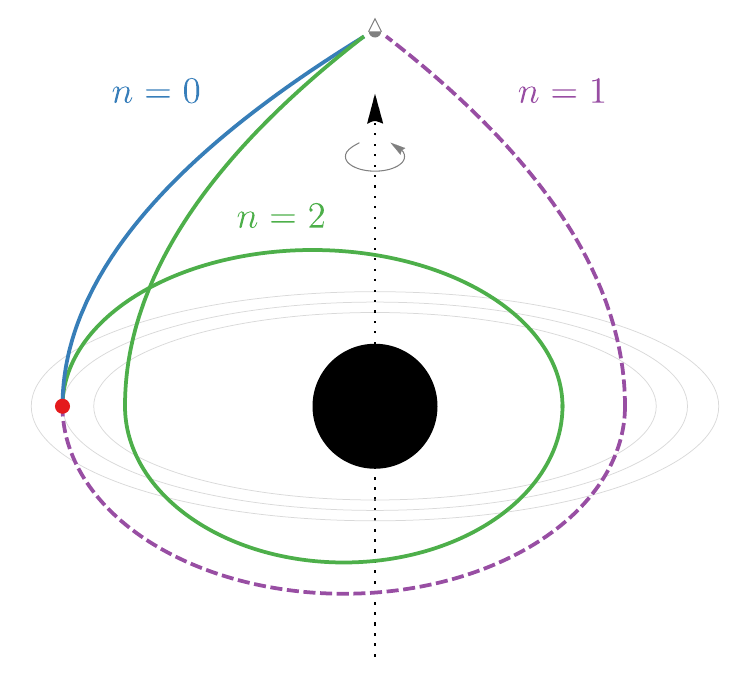}
    \caption{{\bf The photon shell of spherical bound orbits.}
    Left (reproduced from Fig.~3(e) of Teo \cite{Teo2021}): Example of such an orbit.
    Middle (adapted from Fig.~2 of Johnson et al \cite{JohnsonLupsasca2020}): The collection of such orbits constitutes the black hole ``photon shell'': the region of spacetime outside the event horizon whose geometry determines the structure of the photon ring.
    Each orbital radius $\tilde{r}$ has a different maximal polar angle $\theta_+$ of libration.
    Right (reproduced from Fig.~7 of Gralla \& Lupsasca \cite{GrallaLupsasca2020a}): A single point source outside the horizon produces a sequence of images labeled by the number of half-orbits $n$ executed around the black hole by the corresponding light ray, leading to self-similarity within the photon ring (Fig.~\ref{fig:PhotonRingStack}).
    }
	\label{fig:PhotonShell}
\end{figure}

A black hole is a region of spacetime in which gravity is so strong that nothing can escape from it---not even light, no matter what direction it is shone in.
The surface enclosing this region of no return is the event horizon (black circle in middle panel of Fig.~\ref{fig:PhotonShell}).\footnote{Here, we restrict our attention to the astrophysically relevant Kerr black hole, whose spacetime geometry is fully determined by the mass $M$ and angular momentum $J=aM$ around the spin axis.
We always work in Boyer-Lindquist coordinates, in which the (outer) event horizon is located at radius $r_+=M+\sqrt{M^2-a^2}$.}
Outside the horizon, light can in principle still escape along some directions.
However, there is a region known as the ``photon shell'' (orange region in middle panel of Fig.~\ref{fig:PhotonShell}) in which gravity is still strong enough to deflect light rays to such an extreme degree that they remain trapped within the shell, orbiting at fixed radius $\tilde{r}$.

This light trapping phenomenon requires a delicate balance between the momentum of a photon and and the gravitational pull of the Kerr black hole.
This equilibrium can only be reached for certain radii $\tilde{r}_-\le\tilde{r}\le\tilde{r}_+$, where the outer-and-innermost radii $\tilde{r}_\pm$ delineate the edges of the photon shell:
\begin{align}
    \tilde{r}_\pm=2M\pa{1+\cos\br{\frac{2}{3}\arccos\pa{\pm\frac{a}{M}}}}.
\end{align}
In the non-rotating (Schwarzschild) limit $a\to0$, these edges coalesce to $\tilde{r}_\pm\to3M$ and the photon shell shrinks to a single ``photon sphere'' slightly outside the event horizon at radius $r_+=2M$.
As the black hole spin increases towards its maximum $a\to M$, the shell thickens until it attains its maximal range $[\tilde{r}_-,\tilde{r}_+]=[M,4M]$.

Bound orbits within the photon shell wind around the black hole in the azimuthal direction $\phi$, but are not free to explore every polar angle $\theta$.
Instead, as illustrated in the middle panel of Fig.~\ref{fig:PhotonShell}, these orbits ``bounce'' up and down between polar angles $0\le\theta_-\le\theta\le\theta_+\le\pi$ with
\begin{align}
   \theta_\pm(\tilde{r})=\arccos\pa{\mp\sqrt{u_+}},\quad
   u_\pm=\frac{\tilde{r}\br{-\tilde{r}^3+3M^2\tilde{r}-2a^2M\pm2\sqrt{M\pa{\tilde{r}^2-2M\tilde{r}+a^2}\pa{2\tilde{r}^3-3M\tilde{r}^2+a^2M}}}}{a^2(\tilde{r}-M)^2}.
\end{align}
The left panel of Fig.~\ref{fig:PhotonShell} illustrates such an orbit.
Throughout this paper, we refer to a period of the polar motion as an ``orbit,'' such that one ``half-orbit'' refers to the path of a photon as it travels from $\theta_-$ to $\theta_+$, or vice versa.

The properties of null geodesics bound in the photon shell of a Kerr black hole were first investigated in 1973 by Bardeen \cite{Bardeen1973} and the bound orbits themselves where analyzed in detail by Teo \cite{Teo2003}.
Leo Stein has produced a compelling, interactive visualization of these orbits, which is available on his website \cite{SteinOrbits}.
The relevance of the photon shell to black hole images and the photon ring was clarified in Johnson et al \cite{JohnsonLupsasca2020}, as we now review.

\subsection{The critical curve, (nearly) bound orbits and their critical parameters}
\label{subsec:Criticality}

\begin{figure}[t]
    \centering
    \includegraphics[width=\textwidth]{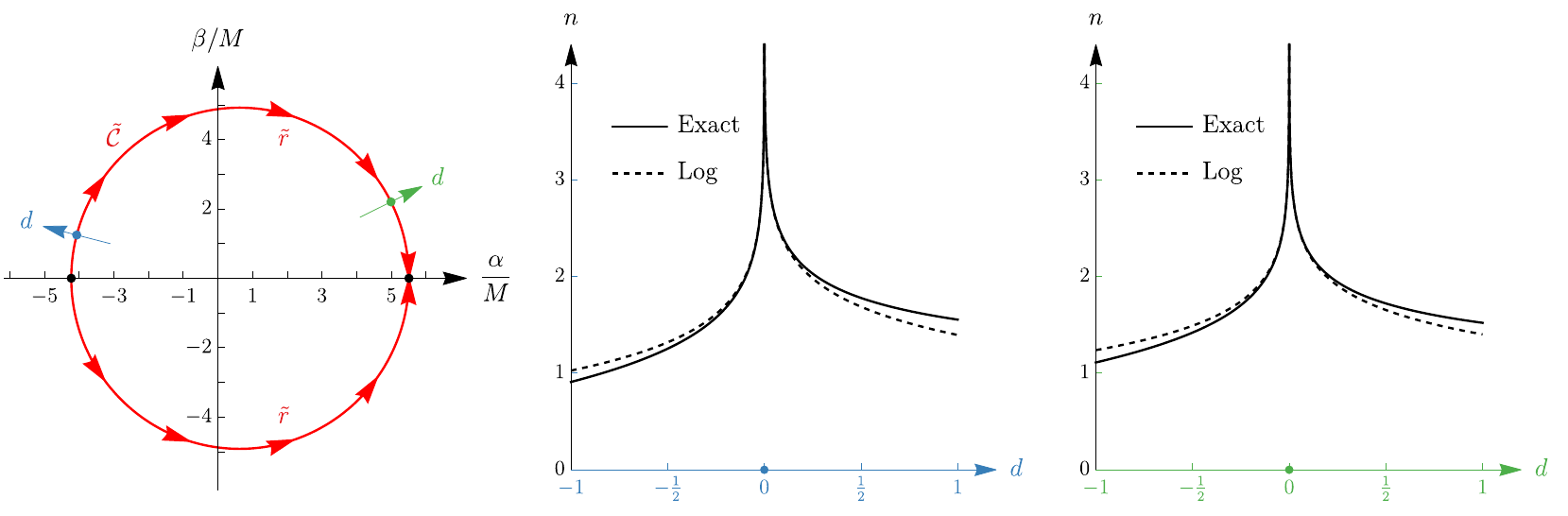}
    \caption{{\bf The critical curve is the image of the photon shell} (adapted from Fig.~3 of Gralla \& Lupsasca \cite{GrallaLupsasca2020a}).
    Left: The image plane of an observer at inclination $\theta_{\rm o}=17^\circ$ from a Kerr black hole with spin $94\%$, parameterized by Bardeen's Cartesian coordinates $(\alpha,\beta)$.
    The critical curve \textcolor{red}{$\tilde{\mathcal{C}}$ (red)} delineates the region of photon capture from that of photon escape and corresponds to light rays that are asymptotically bound in the photon shell.
    Each half of the curve (upper or lower) is parameterized by the orbital radius $\tilde{r}$.
    Middle and right: Fractional number of half-orbits $n$ as a function of (signed) perpendicular distance $d$ from the critical curve \textcolor{red}{$\tilde{\mathcal{C}}$} on the image plane of the distant observer on the left.
    The \textcolor{blue}{middle (blue)} and \textcolor{green}{right (green)} cuts correspond to \textcolor{blue}{$\tilde{r}=2.1M$} and \textcolor{green}{$\tilde{r}=2.9M$}, respectively.
    }
	\label{fig:CriticalCurve}
\end{figure}

In order to describe black hole images, one must first set up a coordinate system on the image plane of a distant observer located at a large distance $r_{\rm o}\gg M$ and inclination $\theta_{\rm o}$ from the hole.
Bardeen \cite{Bardeen1973} introduced coordinates such that a photon reaching such an observer with four-momentum $p^\mu$ appears at Cartesian position $(\alpha,\beta)$ with
\begin{align}
    \alpha=\frac{p_\phi}{p_t\sin{\theta_{\rm o}}},\quad
    \beta=-\frac{p_\theta}{p_t}.
\end{align}
These coordinates are such that the origin corresponds to the ``line of sight to the center of the black hole'' and the $\beta$ axis is the projection of the black hole spin axis onto the plane perpendicular to this line of sight (see, e.g., Fig.~6 and App.~E of Gralla et al \cite{Gralla2018} for details).
We illustrate this coordinate system in the left panel of Fig.~\ref{fig:CriticalCurve}.

Photons aimed backwards into the geometry from the image plane can meet different fates.
If they are aimed towards the black hole (close to the origin), then they eventually cross its event horizon, never to be seen again.
If they are aimed far from the black hole (far away from the origin), then they will experience some deflection due to the gravitational of the black hole but will eventually encounter a radial turning point and return to infinity.
The boundary curve in the image plane delineating the region of photon capture from that of photon escape is known as the ``critical curve'' $\tilde{\mathcal{C}}$ (shown in red in the left panel of Fig.~\ref{fig:CriticalCurve}).
It was first derived by Bardeen \cite{Bardeen1973}.

Photons aimed in a direction corresponding to a point on this critical curve fall towards the black hole, but can neither enter its event horizon nor turn back towards infinity; instead, they asymptote to a bound orbit in the photon shell.
In fact, $\tilde{\mathcal{C}}$ is parameterized by the orbital radius $\tilde{r}$ where such a photon ends up trapped:
\begin{gather}
    \label{eq:CriticalCurve}
    \tilde{\mathcal{C}}=\cu{\pa{\tilde{\alpha}(\tilde{r}),\tilde{\beta}(\tilde{r})}\Big|\tilde{r}\in[\tilde{r}_-,\tilde{r}_+]},\\
    \tilde{\alpha}(\tilde{r})=\frac{\tilde{r}^2(\tilde{r}-3M)+a^2(\tilde{r}+M)}{a(\tilde{r}-M)\sin{\theta_{\rm o}}},\quad
    \tilde{\beta}(\tilde{r})=\pm\sqrt{\frac{\tilde{r}^3\br{4Ma^2-\tilde{r}(\tilde{r}-3M)^2}}{a^2(\tilde{r}-M)^2}-\br{\tilde{\alpha}^2(\tilde{r})-a^2}\cos^2{\theta_{\rm o}}}.
\end{gather}
In other words, the critical curve is the image of the photon shell.
The choice of sign in $\tilde{\beta}(\tilde{r})$ determines whether one traces the upper or lower half of the critical curve.
Physically, this degeneracy stems from the fact that an asymptotically bound photon can enter its orbital shell on either an upwards or downwards bounce.

The critical curve plays a key role in the study of (nearly) bound photons, which are crucial for understanding black hole images.
We stress, however, that $\tilde{\mathcal{C}}$ is a mathematical curve, and is therefore not in itself observable.
It is also an infinitely thin curve.
Physically, this corresponds to the fact that light trapping within the photon shell is unstable, so one must aim a photon infinitely precisely for it to keep orbiting forever (albeit unstably).

Nevertheless, a photon that is aimed very near---but not on---the critical curve will spend a very long time orbiting around the black hole within its photon shell, before eventually falling into the event horizon (if it was aimed slightly inside of $\tilde{\mathcal{C}}$) or bouncing back to radial infinity (if it was aimed slightly outside of $\tilde{\mathcal{C}}$).
Thus, such ``near-critical'' photons follow nearly bound orbits.

More precisely, consider a photon aimed at a (signed) perpendicular distance $d$ from a point with coordinates $\tilde{\alpha}(\tilde{r})$ and $\tilde{\beta}(\tilde{r})$ on the critical curve, as illustrated in Fig.~\ref{fig:CriticalCurve}.
Gralla \& Lupsasca \cite{GrallaLupsasca2020a} proved that such a photon skirts the nearby photon orbit at radius $\tilde{r}$ in the photon shell, executing around the black hole a fractional number\footnote{See their Eq.~(34) for a precise definition of the fractional orbit number and their Eqs.~(74)--(76) for a closer logarithmic approximation, shown in the middle and right panels of Fig.~\ref{fig:CriticalCurve} (dashed lines) against the exact analytic result (solid curves).}
\begin{align}
    \label{eq:LogDivergence}
    n\approx-\frac{1}{\gamma(\tilde{r})}\log\ab{d}
\end{align}
of half-orbits along its full trajectory.
The coefficient $\gamma(\tilde{r})$ controlling this logarithmic divergence is given by \cite{JohnsonLupsasca2020}
\begin{align}
    \gamma(\tilde{r})=\frac{4}{a}\sqrt{\tilde{r}^2-\frac{M\tilde{r}\pa{\tilde{r}^2-2M\tilde{r}+a^2}}{(\tilde{r}-M)^2}}\int_0^1\frac{\ed t}{\sqrt{\pa{1-t^2}\pa{u_+t^2-u_-}}}.
\end{align}
Physically, $\gamma(\tilde{r})$ corresponds to the Lyapunov exponent governing the orbital instability of a photon bound at radius $\tilde{r}$ in the photon shell: if such a photon is pushed off its orbital shell by some small amount $\delta r_0=r-\tilde{r}$, with $\ab{\delta r_0}\ll M$, then after $n$ half-orbits, its radial position $\delta r_n$ will have diverged exponentially as
\begin{align}
    \delta r_n\approx\delta r_0e^{\gamma(\tilde{r})n}.
\end{align}
Visualizations of this unstable behavior are presented in a companion paper by Galison et al \cite{Galison2024}.

The Lyapunov exponent $\gamma(\tilde{r})$ depends only on the orbital radius and the mass and spin parameters of the black hole.
This function can be viewed a fundamental property of black holes predicted by general relativity.
It is often referred to as a ``critical exponent'' for two reasons.
First, the Lyapunov exponent $\gamma$ is a function of the parameter $\tilde{r}$ along the critical curve \eqref{eq:CriticalCurve}, and in this sense can be viewed as a critical parameter along $\tilde{\mathcal{C}}$.
Second, when viewed as a region of (photon) {\it phase space}, rather than as a region of {\it spacetime}, the photon shell emerges as a special locus displaying conformal symmetry.\cite{Hadar2022,Kapec2023}
Emergent conformal symmetry is the hallmark of a critical point, and part of the modern definition of critical phenomena.

Two other critical parameters $\delta(\tilde{r})$ and $\tau(\tilde{r})$, respectively defined as the azimuth swept $\Delta\phi$ and time lapse $\Delta t$ incurred by a bound photon over the course of one half-orbit, complete the analytic theory of the photon ring \cite{GrallaLupsasca2020a}.

A single luminous point source outside the black hole gives rise to a sequence of images $(\alpha_n,\beta_n)$ in the sky of a distant observer, with each image arising from a strongly bent ray that has circulated some number of times $n$ around the black hole (see right panel of Fig.~\ref{fig:CriticalCurve}).\footnote{As the half-orbit number $n$ grows large, there can be multiple images with the same $n$.
Therefore, to uniquely label each image, one needs an additional integer $m$ counting the number of azimuthal windings along each photon's trajectory.}
According to Eq.~\eqref{eq:LogDivergence}, if the $n^\text{th}$ image of the source appears at some perpendicular distance $d_n$ from $\tilde{\mathcal{C}}$, then the next image must appear exponentially closer, at a distance
\begin{align}
    d_{n+1}\approx e^{-\gamma}d_n
    \approx e^{-n\gamma}d_1.
\end{align}
Thus, successive images of a point source accumulate near the critical curve, growing closer to it exponentially fast in $n$.
In addition, this relation implies that if the source is not a point, but rather has a small extent, then its successive images are also exponentially demagnified, with the $n^\text{th}$ image demagnified by a factor $\sim e^{-\gamma n}$.\cite{Luminet1979,Gralla2019}.

Moreover, if the source is blinking, then successive flashes ought to be observed with a time delay of $\Delta t\approx\tau$, such that the $n^\text{th}$ ``echo'' appears $\Delta t\sim n\tau$ after the first flash.
Gralla \& Lupsasca \cite{GrallaLupsasca2020a} make these relations precise and show that they are in fact asymptotically exact (as $n\to\infty$) in the simplest case of an equatorial source viewed from the spin axis, depicted in the right panel of Fig.~\ref{fig:CriticalCurve}.
In that case, successive images appear rotated relative to each by a polar image angle $\Delta\varphi=\delta$, which exactly matches the additional azimuth swept by the successive photons on their way to the observer; see their Fig.~8 for an illustration of this lensing behavior.

The upshot is that $\gamma$, $\delta$ and $\tau$ completely control the lensing behavior of Kerr black holes in the regime of large $n$.
These spin-dependent critical parameters are new fundamental quantities that encode {\it universal} (i.e., matter-independent) features of general relativity and could in principle be observed with future detectors, either by comparing successive images of a source, or by extracting them from the characteristic correlation structure produced in time-dependent images by the multiple images of lensed fluctuations \cite{Wong2021,Hadar2021}.

A measurement of $\gamma$, $\delta$ or $\tau$ could also provide a test of the Kerr hypothesis, as these critical parameters are predicted to take different values in other black hole spacetimes \cite{Wielgus2021,Staelens2023}.

\subsection{The photon ring and ``black hole shadow''}

\begin{figure}[t]
    \centering
    \includegraphics[width=0.5\textwidth]{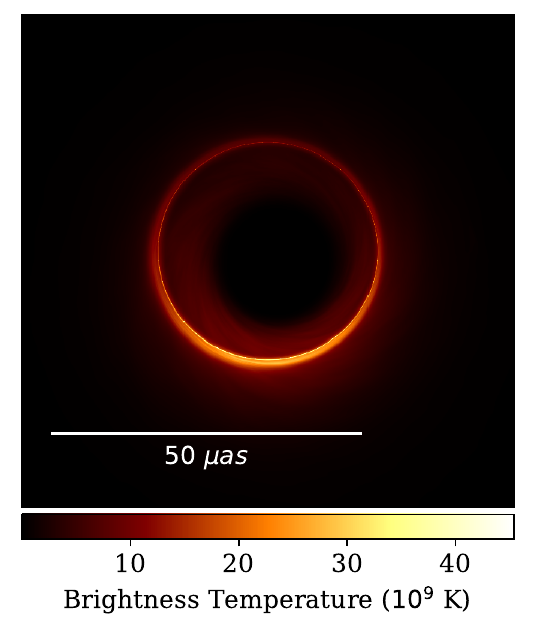}
    \includegraphics[width=0.425\textwidth]{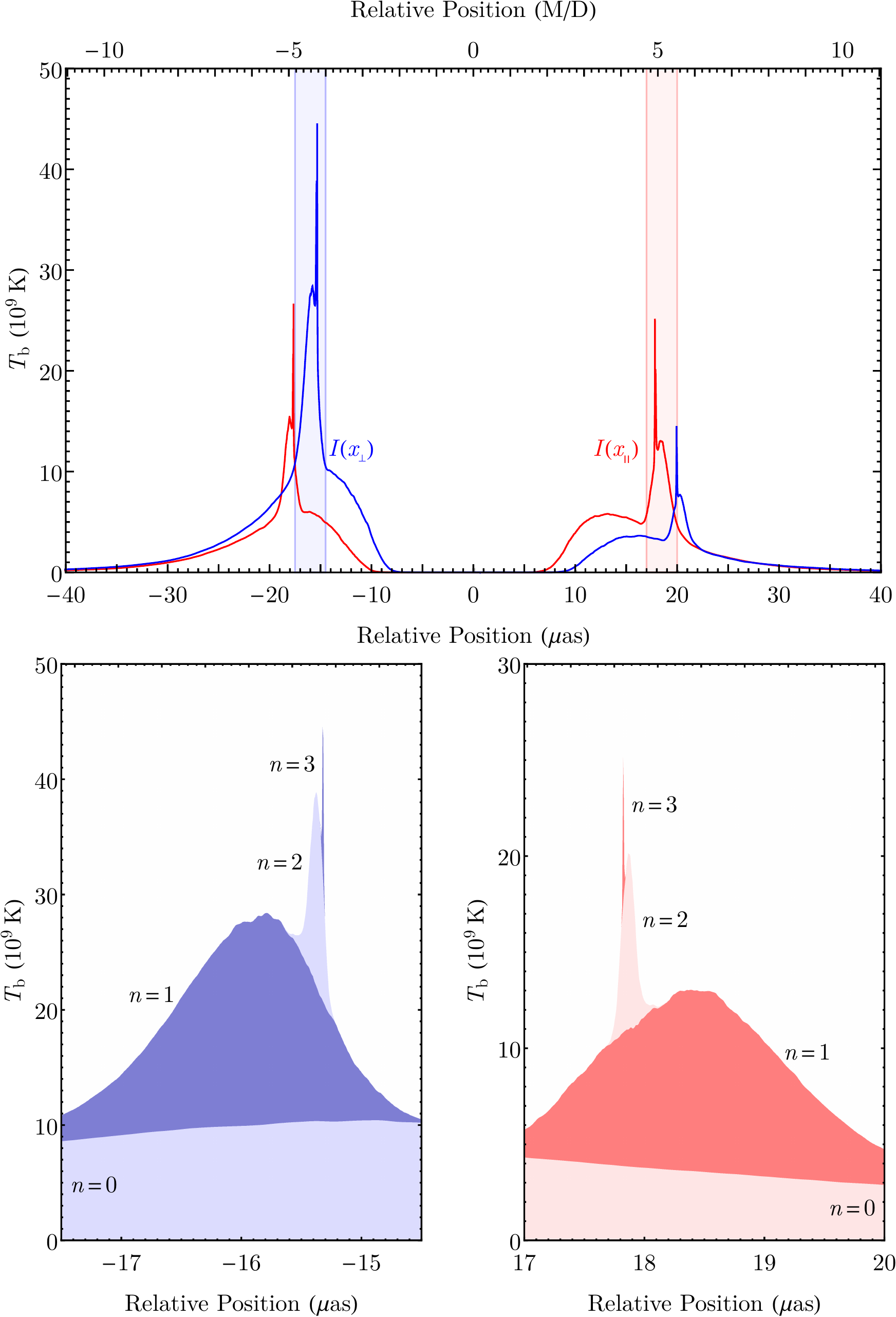}
    \caption{{\bf Time-averaged image of a GRMHD simulation of the supermassive black hole M87* compatible with the 2017 EHT observations} (reproduced from Figs.~1 and 3 of Johnson et al \cite{JohnsonLupsasca2020}).
    Left: Image of a GRMHD flow accreting onto a black hole with spin 94\% viewed from an inclination of $\theta_{\rm o}=163^\circ$.
    The time average is taken over 100 snapshots produced from uniformly spaced samples over a time range of 1,000$M$, where $M\sim9$ hours for M87*.
    Top right: Cross sections of the image intensity.
    The blue/red curves show the intensity along cuts perpendicular/parallel to the projected spin axis $\beta$.
    Bottom right: Decomposition of the intensity peaks corresponding to the photon ring into subrings labeled by photon half-orbit number $n$.
    The rings converge to the shape \eqref{eq:CriticalCurve} of the Kerr critical curve.
    }
	\label{fig:RadialProfile}
\end{figure}

\begin{figure}[t]
    \centering
    \includegraphics[width=.7\textwidth]{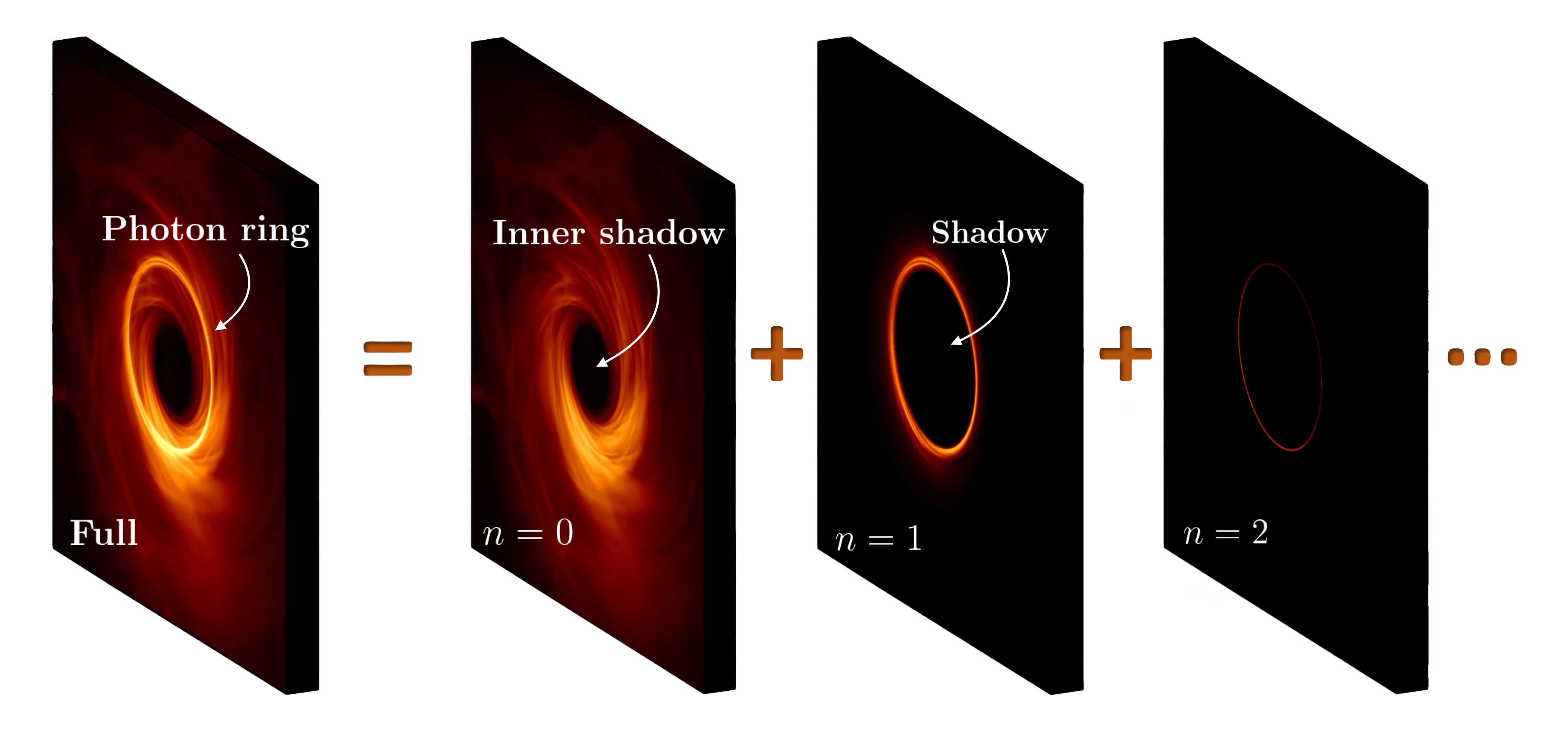}
    \includegraphics[width=.28\textwidth]{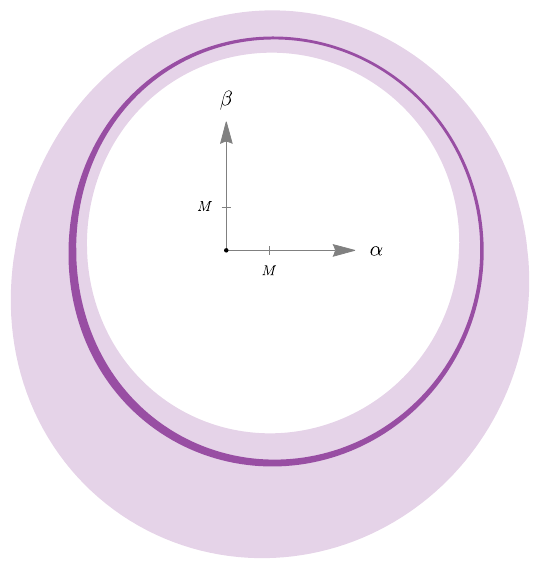}
    \caption{
    {\bf Left: Prominent features in black hole images.}
    Snapshots of a black hole display distinctive features, including a narrow photon ring composed of a stack of increasingly sharp subrings.
    Each subring $n$ is produced by photons that orbited around the black hole $n/2$ times before reaching the observer.
    These subrings combine to give the full image.
    They also provide precise insights about the underlying spacetime via their size, shape, and self-similar structure.
    Other prominent features that carry information about the black hole include its ``shadow'' \cite{Falcke2000} and ``inner shadow'' \cite{Chael2021} but these also depend on details of the underlying emission.
    {\bf Right: Kerr lensing bands.}
    The \textcolor{lightpurple}{$n=1$} and \textcolor{darkpurple}{$n=2$} lensing bands (in \textcolor{lightpurple}{light purple} and \textcolor{darkpurple}{dark purple}, respectively) in the image plane of an observer at an inclination $\theta_{\rm o}=30^\circ$ from a Kerr black hole with spin $99\%$ (reproduced from Fig.~3 of Gralla et al \cite{GLM2020}).
    Each subring is constrained (by lensing considerations alone) to lie entirely within its lensing band \cite{Paugnat2022,AART}.}
    \label{fig:PhotonRingStack}
\end{figure}

While point sources are useful for developing intuition about black hole lensing, to interpret the EHT observations of M87* and Sgr\,A*, one must instead consider emission from more realistic sources around a black hole, such as an accretion flow of hot, radiating gas.
Such flows can be modeled with state-of-the-art GRMHD simulations \cite{EHT2019e}, and their emission numerically ray traced using the equations of radiative transfer \cite{Dexter2016}.

A key idea for understanding the photon ring effect is the following observation: light rays that spend more time traversing the radiating source around the black hole tend to collect more photons along their path.
Therefore, if one shoots multiple rays backwards into the geometry from the image plane of a distant observer, then those rays that spend more time orbiting in the emission region around the black hole will be loaded with more photons from the radiating matter, and the intensity observed at the corresponding image-plane position will be greater.
As a result, one can expect the observed intensity $I_{\rm o}$ of image pixels $(\alpha,\beta)$ to grow rapidly near the critical curve $\tilde{\mathcal{C}}$, as the path length of the corresponding rays diverges logarithmically according to Eq.~\eqref{eq:LogDivergence},
\begin{align}
    \label{eq:IntensityPeak}
    I_{\rm o}(\alpha,\beta)\sim n
    \approx-\frac{1}{\gamma(\tilde{r})}\log\ab{d}.
\end{align}
In other words, a ray that orbits $n$ times through the emission region can collect roughly $\sim n$ times as many photons as are loaded onto a ray that crosses the region only once.
Of course, this argument is not quantitatively correct, as it is based purely on the diverging path length of near-critical rays and neglects relativistic effects (such as Doppler beaming) as well as the details of radiative transfer, which affect the precise number of photons loaded onto each ray.
Nevertheless, these complications do not qualitatively modify the prediction \eqref{eq:IntensityPeak}, which is indeed observed across many models.

The photon ring is then defined as the bump in the observed intensity that appears near the critical curve and contains the logarithmic divergence \eqref{eq:IntensityPeak}.\cite{JohnsonLupsasca2020}
No such infinity is ever actually observed in realistic models, since absorption effects naturally cut off the divergence, as photons that spend a long time orbiting around the black hole after their emission have an increasing chance of being reabsorbed.

We now briefly describe two broad classes of emission models displaying photon rings with different structures.

First, if a black hole is fully immersed within a completely spherical accretion flow, then the observed intensity near the critical curve really does grow with the path length of rays and therefore increases continuously towards $\tilde{\mathcal{C}}$ as in the middle and right panels of Fig.~\eqref{fig:CriticalCurve}.
Such a configuration was first studied by Falcke et al \cite{Falcke2000} and recently revisited by Narayan et al \cite{Narayan2019}, who showed that the intensity displays the scaling behavior \eqref{eq:IntensityPeak} regardless of the details of the flow.
In all such models, the emission produces a bright ring whose intensity spikes precisely on the critical curve $\tilde{\mathcal{C}}$, while the interior of $\tilde{\mathcal{C}}$---which corresponds to the black hole, that is, light rays that end on the horizon---remains dark.
This distinctive feature---a brightness deficit inside $\tilde{\mathcal{C}}$---is the ``black hole shadow''.

We stress that, while the critical curve $\tilde{\mathcal{C}}$ {\it is} directly observable in the context of spherical accretion models, this is typically not the case.
Indeed, such ``highly fine-tuned'' scenarios are  disfavored by the 2017 EHT observations, which are more compatible with accretion flows in ``magnetically arrested disk'' (MAD) states.\cite{EHT2019e}.
Generically, such GRMHD flows form geometrically thick disks with mostly near-equatorial emission, as illustrated in Fig.~4 of the 2019 EHT paper V interpreting the 2017 data.\cite{EHT2019e}.
The time-averaged image of such a flow is shown in the left panel of Fig.~\ref{fig:RadialProfile}, with the corresponding intensity cross sections plotted in the right panel thereof.

Since the emission region in such models is not entirely continuous (because it has ``gaps'' above and below the disk), the image intensity profile does not display the continuously growing behavior illustrated in Fig.~\ref{fig:CriticalCurve}.
Instead, the right panel of Fig.~\ref{fig:RadialProfile} shows a series of subrings within the photon ring, which each successive subring corresponding to a full image of the direct emission lensed by an additional half-orbit around the black hole.
The resulting intensity profile displays a series of ``bites'' taken out of the logarithmic profile \eqref{eq:IntensityPeak}, with the $n^\text{th}$ bite corresponding to the intensity that would be contributed by the $n^\text{th}$ image of the gap in the emission region, were this gap to be filled.
Thus, the image of a black hole surrounded by a near-equatorial (geometrically thin or thick) disk displays a photon ring with a characteristic subring structure, illustrated in the left panel of Fig.~\ref{fig:PhotonRingStack}.

\begin{figure}[t]
    \centering
    \includegraphics[width=.75\textwidth]{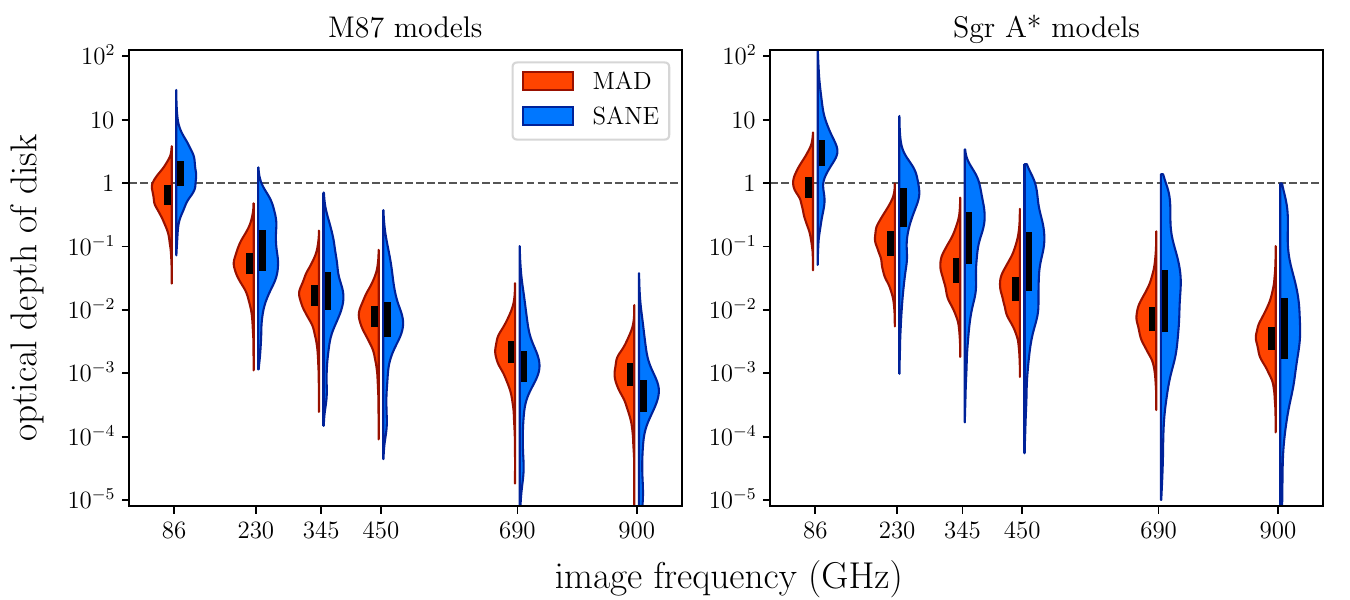}
    \caption{
    {\bf Optical depth as a function of observing frequency in GRMHD simulations of accretion flows onto the supermassive black holes M87* and Sgr\,A*.}
    Soon-to-be-published plot provided by G.~N.~Wong \cite{Wong2024}.
    These simulations predict that the $n=1$ and $n=2$ rings are almost always visible at 230\,GHz and at 345\,GHz, respectively.
    This is the case for both ``magnetically arrested disk'' (MAD) and ``standard and normal evolution'' (SANE) states.}
    \label{fig:OpticalDepth}
\end{figure}

According to the discussion in Sec.~\ref{subsec:Criticality}, this substructure is universal (i.e., weakly dependent on the details of the source), since the subring images are related to each other by a successive exponential demagnification, rotation and time delay controlled by the GR-predicted critical parameters $\gamma$, $\delta$, and $\tau$, respectively.
Essentially, the only two ingredients required to observe such a sequence of subrings are: 1) an emission region with ``gaps'' and 2) an optically thin source, such that photons can execute a large number $n$ of half-orbits without being reabsorbed.
Optical depth increases with observing frequency, and existing EHT images indicate that its 230\,GHz observations are already in the optically thin regime where at least the first $n=1$ ring is visible.
Semi-analytic thick disk models \cite{Vincent2022} as well as GRMHD simulations (see Fig.~\ref{fig:OpticalDepth}) suggest that the $n=1$ subring always ought to be present at 230\,GHz, with the $n\le2$ subrings almost always visible at the target BHEX frequency of 300\,GHz.

As shown in Fig.~\ref{fig:PhotonRingStack}, the higher-order ($n>0$) layers of the full image display a photon subring encircling a brightness deficit---the ``shadow'' effect---but the $n=0$ layer generically displays emission within the subring.
This emission typically extends down to the event horizon in the equatorial plane of the black hole, giving rise to a brightness deficit within the direct ($n=0$) image of the equatorial event horizon.
This image feature is known as the ``inner shadow'' and also encodes properties of the spacetime geometry.\cite{Chael2021}

\subsection{Lensing bands and black hole parameter extraction}
\label{subsec:LensingBands}

As can be seen from the image layers indexed by half-orbit number $n$ in Figs.~\ref{fig:RadialProfile} (right panel) and \ref{fig:PhotonRingStack} (left panel), the subring images that make up the photon ring stack on top of one another.
According to the lensing behavior described in Sec.~\ref{subsec:Criticality}, these successive lensed images of the main emission are exponentially demagnified by $\sim e^{-\gamma}$.
Moreover, by Eq.~\eqref{eq:LogDivergence}, they must also converge to the critical curve $\tilde{\mathcal{C}}$ exponentially fast in $n$.
These two key properties follow purely from the structure of the photon shell and the spacetime geometry of the black hole.

To make these statements more precise, it is useful to introduce a notion of ``lensing band'' in the image plane of a distant observer \cite{GLM2020,Paugnat2022,AART}.
The $n^\text{th}$ (equatorial) lensing band $\mathcal{L}_n$ is the region of the image plane corresponding to those rays whose maximal extension in the spacetime geometry crosses the equatorial plane $n+1$ times before either terminating on the event horizon (if the ray lies at a negative perpendicular distance $d<0$ in the interior of $\tilde{\mathcal{C}}$) or returning to infinity (if it lies at a positive distance $d>0$ in the exterior of $\tilde{\mathcal{C}}$).
According to Eq.~\eqref{eq:LogDivergence}, a near-critical ray can skirt its nearby bound orbit in the photon shell $n$ times with either sign of $d$.
Therefore, it follows that each lensing band straddles the critical curve: $\tilde{\mathcal{C}}\subset\mathcal{L}_n$.
Moreover, a ray that crosses the equator $k$ times must evidently lie within all the lensing bands $\mathcal{L}_n$ with $n<k$.
As such, the bands form a nested structure
\begin{align}
    \mathcal{L}_0\supset\mathcal{L}_1\supset\mathcal{L}_2\supset\ldots\supset\tilde{\mathcal{C}}=\mathcal{L}_\infty\equiv\lim_{n\to\infty}\mathcal{L}_n,
\end{align}
We show an example of the $n=1$ and $n=2$ lensing bands in the right panel of Fig.~\ref{fig:PhotonRingStack}.

The usefulness of this definition lies in the observation that the $n^\text{th}$ subring image of an (equatorial) disk must necessarily lie within the $n^\text{th}$ lensing band, {\it regardless of the details of the source}.\footnote{Given that GRMHD models favor near-equatorial emission, it is particularly convenient to consider equatorial lensing bands, but when needed, their definition can easily be extended to cones comprising other source inclinations $\theta_{\rm s}\neq90^\circ$.}
At finite $n$, the $n^\text{th}$ subring of the (equatorial) source is some thin ring whose precise appearance {\it does} depend to some extent on the details of the source, but these details become progressively irrelevant (exponentially fast) at large $n$.

Indeed, for large enough $n$, the $n^\text{th}$ lensing band is so narrow as to effectively be a thin plane curve $\mathcal{L}_n\approx\mathcal{C}_n$, with all possible emission profiles producing effectively the same shape $\mathcal{C}_n$ (up to exponentially small differences suppressed by $e^{-n\gamma}$).
Moreover, up to exponentially small corrections (also suppressed by $e^{-n\gamma}$), this shape is effectively identical to that of the critical curve, which can be viewed as the limiting ``$n\to\infty$'' photon subring:
\begin{align}
    \label{eq:InfiniteSubring}
    \tilde{\mathcal{C}}=\mathcal{C}_\infty\equiv\lim_{n\to\infty}\mathcal{C}_n.
\end{align}
In practice, the approximation $\mathcal{L}_n\approx\tilde{\mathcal{C}}$ is already excellent for $n\gtrsim2$ \cite{GLM2020}, and essentially exact (for all intents and purposes) in the ``universal regime'' $n\gtrsim3$ \cite{JohnsonLupsasca2020}.
As such, measuring a photon subring with $n\gtrsim3$ would amount to a measurement of the (astrophysics-independent) critical curve \eqref{eq:CriticalCurve}, and hence of the black hole parameters.

As for the $n=1$ photon ring targeted by future observations with BHEX, it does not yet lie completely within the universal regime of large $n\gtrsim2$.
This is reflected in the non-negligible width of the $n=1$ lensing band $\mathcal{L}_1$, which is of order $\sim M$ for small-to-modest inclinations such as the one shown in the right panel of Fig.~\ref{fig:PhotonRingStack}.
This significant width implies that different emission profiles within the astrophysical source can lead to noticeably different $n=1$ photon ring shapes within $\mathcal{L}_1$.
As such, extracting the black hole mass, spin, and inclination from the shape of the $n=1$ photon ring is not as straightforward as with for the higher-order subrings.

Nevertheless, such a shape measurement would still enable sharp estimates of the black hole parameters.
For instance, the first lensing band $\mathcal{L}_1$ (which contains both the $n=1$ photon ring and the critical curve) has width $w_1\sim M$ and diameter $d\sim 10M$ for most spins and inclinations, that is, a width-to-diameter ratio $w_1/d\sim10\%$.
It follows that a measurement of the $n=1$ photon ring diameter would provide an estimate of the black hole mass-to-distance ratio accurate to within $\sim10\%$ {\it based on purely geometric considerations derived from strong lensing alone}.
Under some (mild) additional assumptions on the source, this precision can rapidly increase.\footnote{By modeling the source with GRMHD, the 2017 EHT observations of M87* (which only resolved the $n=0$ emission) already enabled an estimate of the black hole mass to within $\sim20\%$ \cite{EHT2019a}.
A similar analysis including BHEX observations that resolve the $n=1$ ring is expected to do better.}

To infer black hole the spin $a_*$ and inclination $\theta_{\rm o}$, one must clearly target observations of the ellipticity of the $n=1$ photon ring.
In particular, the ring diameters $d_\parallel$ and $d_\perp$ in the directions parallel and perpendicular to the projected spin axis are particularly sensitive to the black hole spin and inclination, respectively.
Preliminary analyses suggest that one could extract from the $n=1$ ring shape a spin estimate accurate to within $\sim10\%$, with even tighter bounds achievable at high spins (and less sensitive constraints of order $\sim20\%$ at lower spins).

\begin{figure}[t]
    \centering
    \includegraphics[width=.5\textwidth]{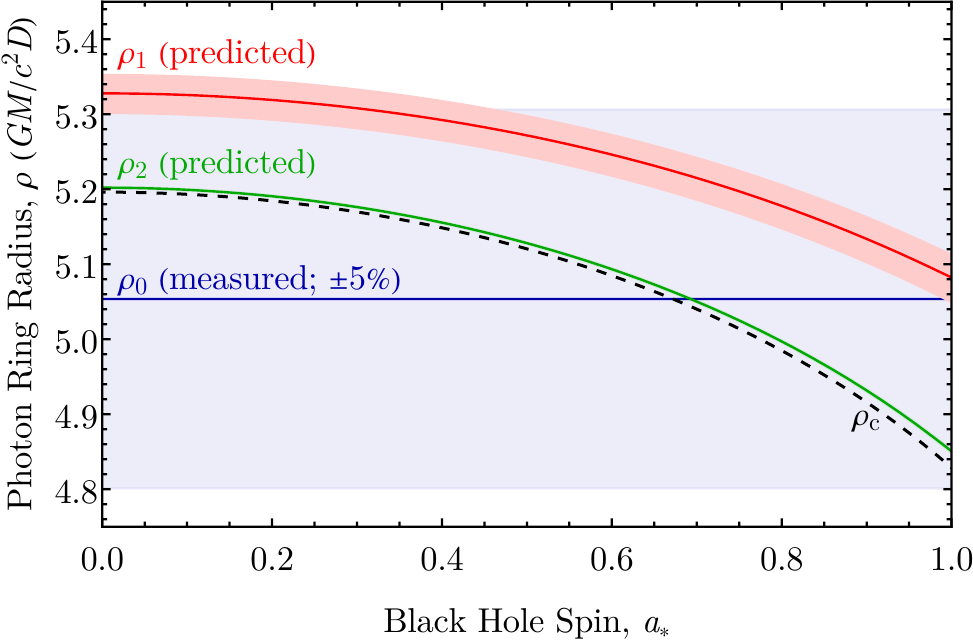}
    \caption{
    {\bf Expected precision of a black hole spin measurement of Sgr A* with BHEX.}
    The relative sizes of successive ring images depend on the black hole spin.
    Since the diameter of the main emission ring has been measured by EHT, a photon ring diameter measurement by BHEX will constrain spin.
    To estimate the precision of such a measurement, we consider a toy model where the emission arises from a thin ring in the equatorial plane of the black hole, which is viewed face-on.
    The measured diameter of the \textcolor{blue}{$n=0$ ring (blue line)} predicts the corresponding diameters of the photon rings (\textcolor{red}{$n=1$ in red}, \textcolor{green}{$n=2$ in green}) as a function of spin.
    The uncertainty in the $n=0$ measurement (blue band) similarly sets the uncertainty in the predictions (red and green bands).
    If limited by this uncertainty, then BHEX can achieve spin measurements with a precision that depends on spin and ranges from $\sim10\%$ (at high spin) to $20\%$ (at low spin).
    }
    \label{fig:SpinMeasurement}
\end{figure}

The most powerful methods for parameter extraction involve both the $n=0$ image of the direction emission and the $n=1$ image of the first photon ring.
Comparing these two images and fitting their various distinctive features shown in Fig.~\ref{fig:PhotonRingStack} (including the inner shadow) provides the most information about the underlying geometry.
As an illustrative example, we show in Fig.~\ref{fig:SpinMeasurement} a simple method that could be used to infer the black hole spin for Sgr\,A* (while M87* spin inference would best be tackled via other approaches).

The method assumes that the black hole mass-to-distance ratio is known to some precision; in Fig.~\ref{fig:SpinMeasurement}, we assume that $M/D$ is known to within 5\%, which is already the case for Sgr\,A* \cite{Do2019,GRAVITY2022}.
In that case, a measurement of the diameter of the $n=0$ ring produced by the direct emission, which can be carried out on Earth baselines (and has been with the EHT \cite{EHT2022a}), can be used to infer an effective radius of emission for the source.
In turn, this leads to a prediction for the diameters of the $n\ge1$ photon rings as (sensitive) functions of the black hole spin.

More precisely, in Fig.~\ref{fig:SpinMeasurement}, we consider a simple toy model consisting of a thin ring of equatorial emission viewed ``face-on'' (that is, from the spin axis, $\theta_{\rm o}=0^\circ$), in which case the equatorial radius $r_{\rm s}$ of the source ring can be inferred in units of $M$ to within $5\%$; the plot assumes an $n=0$ ring radius of $\rho_0\approx5.05M$, corresponding (via the ``just add one'' prescription \cite{GrallaLupsasca2020a}) to a source radius $r_{\rm s}\approx4.05M\pm5\%$ (blue band).
The predicted $n=1$ ring radius $\rho_1$ is then a rather sensitive function of spin, occupying a narrow range (red band) whose uncertainty is demagnified by $\sim e^{-\gamma}$ relative to that of the inferred emission radius (blue band).
If BHEX measurements of the $n=1$ ring radius are sufficiently precise (narrower than the width of the red band), then Fig.~\ref{fig:SpinMeasurement} suggests that they could be used infer the spin of Sgr\,A* to within $20\%$ at low spins, or $10\%$ at high spins.
Simulated interferometric measurements of the $n=1$ ring diameter using space-VLBI can achieve this requisite precision \cite{CardenasAvendano2024}.

This method is not suitable for M87*, given the larger uncertainty in its $M/D$.
However, an angle-dependent photon ring diameter can be measured for M87*, providing shape information that depends on spin.
Work is underway to estimate the precision of a spin measurement in this case.
In particular, fitting methods suited to BHEX data that rely on either geometric modeling or state-of-the-art imaging tools (or a combination thereof) are the subject of active and ongoing research.

\section{The photon ring in the visibility domain}
\label{sec:RingVisibility}

The previous section reviewed the emergence of the photon ring as a prominent feature in black hole images.
However, interferometric arrays---such as the Event Horizon Telescope or its extension to space via BHEX---do not directly sample a black hole image, but rather its radio visibility, which is the Fourier transform of the image,
\begin{align}
	\label{eq:ComplexVisibility}
	V(\bm{u})=\int I_{\rm o}(\bm{x})e^{-2\pi i\bm{u}\cdot\bm{x}}\ed^2\bm{x}.
\end{align}
This formula is the van Cittert-Zernike theorem.
It relates the image intensity $I_{\rm o}(\alpha,\beta)$, parameterized in terms of (dimensionless) angular coordinates on the image plane $\bm{x}\approx(\alpha,\beta)/r_{\rm o}$, to the (generically complex) visibility $V(\bm{u})$ of the source; here, the (dimensionless) vector $\bm{u}=(u,v)$ is a ``baseline'' that corresponds to the distance between telescopes in the array, projected onto the plane perpendicular to the line of sight (the ``baseline'' or ``$uv$'' plane)  and measured in units of the observation wavelength.\cite{Roberts1994}
An interferometric array such as the EHT and its extension to space via BHEX can only sample the radio visibility $V(\bm{u})$ on a sparse set of baselines (e.g., right panel of Fig.~\ref{fig:BHEX}).
Nevertheless, such observations suffice to detect the photon ring and measure its shape \cite{JohnsonLupsasca2020}.

In this section, we review the characteristic interferometric signature of the photon ring and its universal subring structure reviewed in Sec.~\ref{sec:RingImage}.
The main takeaways from this section are:
\begin{itemize}
    \item The photon ring is not only a universal (i.e., matter-independent) feature of sources around a black hole that dominates their appearance in the {\it image domain}, it also produces a strong, universal signature in the {\it Fourier domain} that is directly sampled by a radio interferometer such as the EHT or BHEX \cite{JohnsonLupsasca2020}.
    \item Just as the photon ring has a universal substructure in the image domain consisting of a stack of discrete subrings (Fig.~\ref{fig:PhotonRingStack}), its interferometric signature also displays a universal ``cascading'' structure on very long baselines, with each subring dominating the signal in an annular region \eqref{eq:DominantRing} of the visibility plane \cite{JohnsonLupsasca2020}.
    \item Within the baseline regime dominated by the $n^\text{th}$ subring, the radio visibility \eqref{eq:ComplexVisibility} follows a ``universal'' form \eqref{eq:UniversalVisibility} whose amplitude \eqref{eq:UniversalVisamp} depends only on the angle-dependent ring diameter $d_\varphi^{(n)}$ \cite{Gralla2020,GrallaLupsasca2020c}.
    These diameters must follow a specific functional form \eqref{eq:Circlipse} predicted by GR for the {\it shape} of the photon rings \cite{GLM2020,CardenasAvendano2023}.
    \item An interferometric array with a single orbiting element in space is sufficient to measure the shape of a photon subring: the index $n$ of the subring is determined by the baseline length sampled by the orbiter and the projected diameter $d_\varphi^{(n)}$ at angle $\varphi$ is measurable from angles $(\varphi,\varphi+\pi)$ around the orbit (Fig.~\ref{fig:Diameter}).
\end{itemize}
Here, we explain these statements in the simple context of time-averaged images observed with perfect resolution.
In the next section, we will model snapshot observations of a time-dependent source in the presence of instrument noise, and argue that these conclusions still hold in the realistic context of a space-VLBI mission such as BHEX.

\subsection{Interferometric signature of a thin circular ring}
\label{subsec:DeltaFunctionRing}

\begin{figure}[t]
    \centering
    \includegraphics[width=\textwidth]{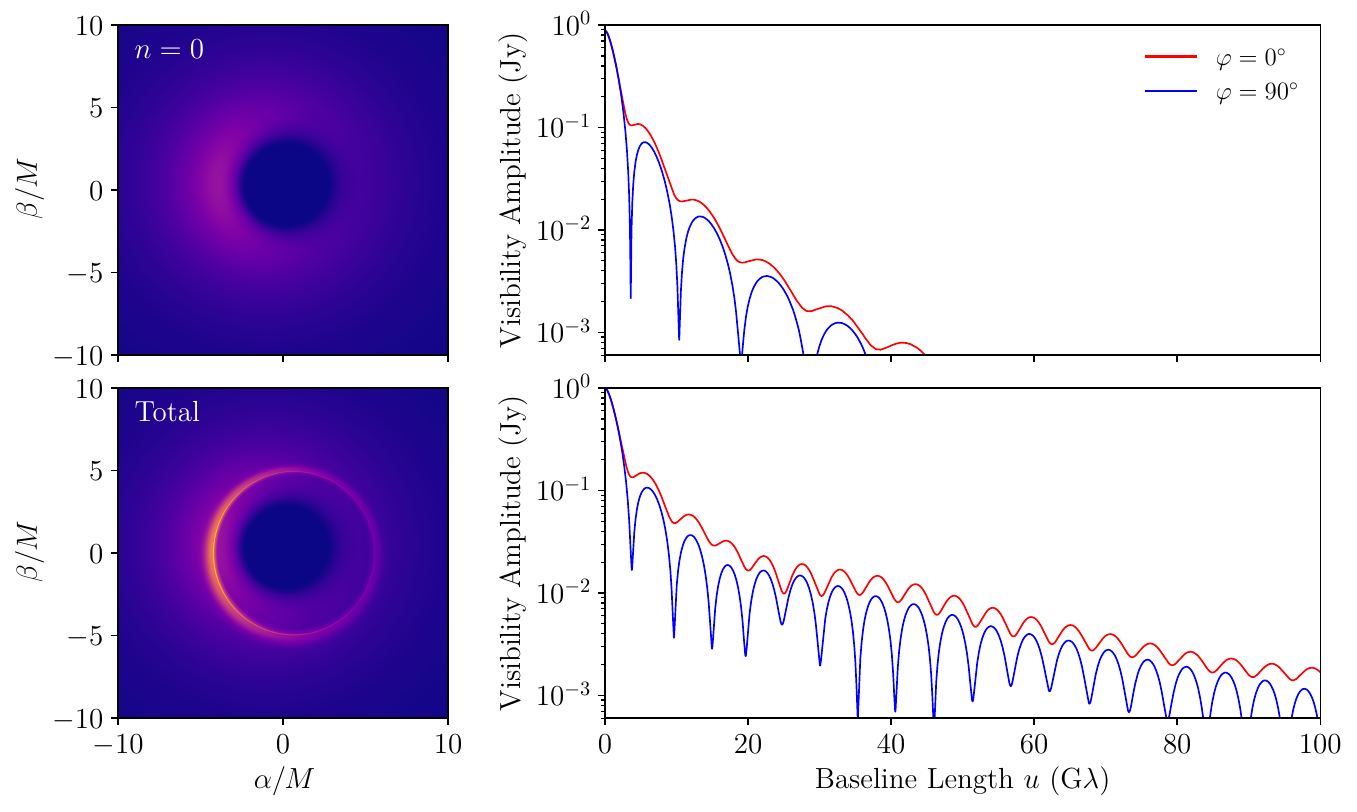}
    \caption{{\bf The black hole photon ring produces a strong interferometric signature in the radio visibility of the source} (reproduced from Fig.~5 of C\'ardenas-Avenda\~no \& Lupsasca \cite{CardenasAvendano2023}).
    Top: Image of an equatorial disk around a black hole with the parameters in Fig.~\ref{fig:RadialProfile}, ray traced with strong lensing turned off.
    Bottom: Same source ray traced with strong lensing turned on.
    The photon ring is the distinctive ``stamp'' added by gravity to the image and its Fourier transform (visibility).
    The thick ring in the direct image produces a rapidly decaying signal, whereas the strongly lensed, thin photon ring produces a clear ringing signature on long baselines to space with a periodicity set by the angle-dependent ring diameter $d_\varphi$ in Fig.~\ref{fig:Diameter}.
    Measuring this periodicity on the space baselines targeted with BHEX will provide a clear photon ring detection and measurement of its shape, which encodes information about the black hole mass and spin.
    }
    \label{fig:GravityStamp}
\end{figure}

Before tackling the interferometric signature of the photon ring, we first examine a warm-up example: the interferometric signature of a perfectly circular and infinitely thin bright ring.
If the ring is centered at the origin, has diameter $d$, and unit flux $V(0)=\int I_{\rm o}(\bm{x})\ed^2\bm{x}=1$, then its image intensity in polar coordinates is
\begin{align}
    \label{eq:InfinitelyThinRing}
    I(\rho,\varphi)=\frac{1}{\pi d}\delta\pa{\rho-\frac{d}{2}}.
\end{align}
The corresponding visibility \eqref{eq:ComplexVisibility} in this case is purely real and given by a Bessel function:
\begin{align}
    \label{eq:RingingSignal}
    V(\bm{u})=J_0(\pi du)
    \stackrel{du>1}{\approx}\frac{\cos\pa{\pi du}+\sin\pa{\pi du}}{\pi\sqrt{du}},
\end{align}
where $u=\ab{\bm{u}}$ is the length of the baseline $\bm{u}$, and in the last line we used an asymptotic expansion of the Bessel function, which is already an excellent approximation by $du=1$.

From this example, we learn that a thin ring image produces a ringing interferometric signature consisting of weakly damped periodic oscillations with a mild power-law falloff and a periodicity set by the ring diameter.
This ringing is only discernible on baselines $u\gtrsim1/d$ that are long enough to resolve the diameter of the ring.

If the image contains other, coarser features that vary over a characteristic length scale $L>d$, then these features are ``resolved out'' (i.e., no longer produce significant power) on baselines of length $u\gtrsim1/L$, leaving the ringing signal \eqref{eq:RingingSignal} to dominate on even longer baselines $u\gtrsim1/d>1/L$.
Given its weak $u^{-1/2}$ power-law falloff, this signal can then persist and remain noticeably strong even on extremely long baselines where traditional VLBI intuition would suggest that no feature of the source can produce significant power.

Next, suppose that the ring has some non-zero width $0<w\ll d$, so that it is no longer infinitely thin, but nonetheless remains ``thin'' in the sense that it has a very small width-to-diameter ratio $0<w/d\ll1$.
Intuitively, on baselines $u\lesssim1/w$ that are not long enough to resolve its width, the interferometer cannot (yet) tell that this ring is not infinitely thin, so its image is indistinguishable from Eq.~\eqref{eq:InfinitelyThinRing}.
As such, in the baseline regime
\begin{align}
    \label{eq:UniversalRegime}
    \frac{1}{d}<u\ll\frac{1}{w},\quad
    \pa{0<\frac{w}{d}\ll1},
\end{align}
which exists because the ring is assumed to be ``thin'' in the sense defined above, the visibility of the ring must still be given by Eq.~\eqref{eq:RingingSignal}.
Only interferometers that can access baselines $u\sim1/w$ can tell that the ring is not actually infinitely thin and observe its power decay rapidly on longer baselines $u\gtrsim1/w$ that resolve it out.
As Fig.~\ref{fig:GravityStamp} shows, to conclusively establish the presence of a thin photon ring, one must access very long baselines that resolve the $n=1$ emission.
For M87* and Sgr\,A*, this will turn out to require space-baselines to BHEX.
(Caveat: the existence of the $n=1$ photon ring could still be inferred from shorter baselines that are not fully dominated by its ringing, where a polarization flip in the photon ring \cite{Himwich2020} may produce a strong polarimetric signature \cite{Palumbo2022}.)

The baseline window \eqref{eq:UniversalRegime} is referred to as the ``universal regime'' because every thin (circular) ring produces the exact same visibility \eqref{eq:RingingSignal} within that range, {\it regardless of the details of its radial profile}.
For explicit examples, see App.~A of C\'ardenas-Avenda\~no \& Lupsasca \cite{CardenasAvendano2023}.
Next, we extend these results to generic (non-circular) rings.

\subsection{Interferometric signature of a generic thin ring}

\begin{figure}[t]
    \centering\includegraphics[width=.32\textwidth]{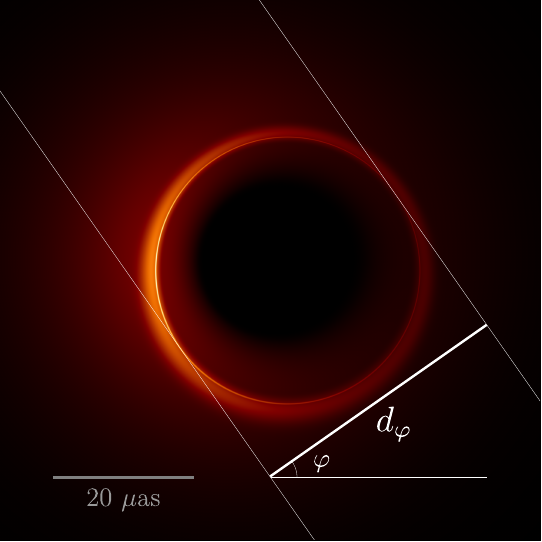}
    \includegraphics[width=.32\textwidth]{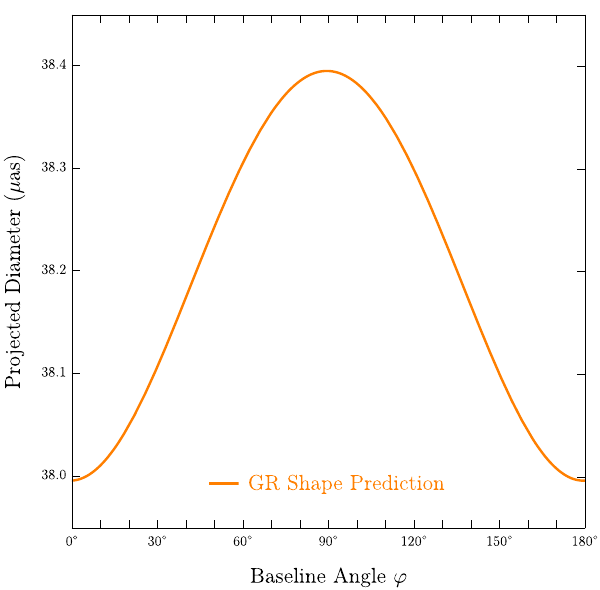}
    \includegraphics[width=.32\textwidth]{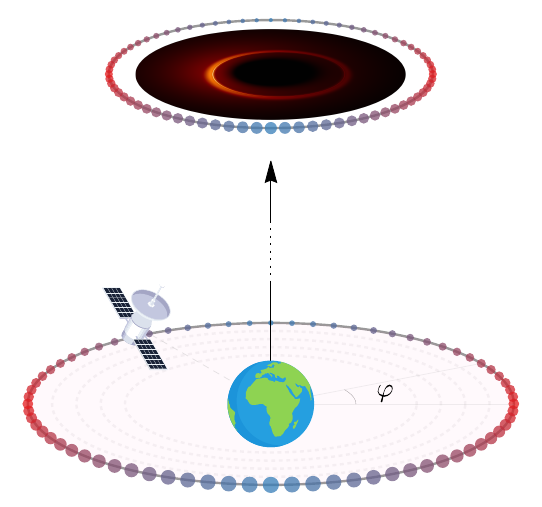}
    \caption{{\bf Measuring the photon ring shape predicted by the Kerr hypothesis in general relativity} (reproduced from Figs.~1 and 2 of Gralla et al \cite{GLM2020}).
    Left: Each photon subring has an angle-dependent diameter $d_\varphi^{(n)}$, shown here for $n=2$.
    Middle: The projected diameters $d_\varphi^{(n)}$ obey a specific GR-predicted functional form \eqref{eq:Circlipse}.
    Right: A satellite orbiting the Earth in the plane perpendicular to the line of sight to the target black hole and observing the source throughout its orbit.
    At angles $(\varphi,\varphi+\pi)$ around the orbit, the baseline to the satellite samples an interferometric ringing signature whose periodicity encodes the diameter $d_\varphi^{(n)}$ of the subring $n$ whose signal dominates the visibility in that region of the baseline plane.
    {\bf A single orbiting space element is enough to measure the photon ring shape.}
    }
    \label{fig:Diameter}
\end{figure}

Consider now a general, convex, image-plane curve $\mathcal{C}$.\footnote{Convexity is not a necessary assumption, but it is sufficient for most photon rings and greatly simplifies the discussion, since non-convex curves require separate normal-angle parameterizations $(x_i(\sigma),y_i(\sigma))$ for each convex segment $i$ \cite{GrallaLupsasca2020c}.}
Suppose we are given some parameterization $(\alpha(\sigma),\beta(\sigma))$ of the curve, and that we wish to compute its visibility \eqref{eq:ComplexVisibility}.
We first compute the normal angle $\varphi$ to the curve,
\begin{align}
    \tan{\varphi(\sigma)}=-\frac{\alpha'(\sigma)}{\beta'(\sigma)}.
\end{align}
Since the curve is by assumption convex, the map $\varphi(\sigma)$ must be invertible.
Therefore, we can define its inverse $\sigma(\varphi)$ and hence obtain the normal-angle parameterization $(\alpha(\varphi),\beta(\varphi))$ of $\mathcal{C}$.
We then define its {\it projected position}
\begin{align}
    f(\varphi)=\alpha(\varphi)\cos{\varphi}+\beta(\varphi)\sin{\varphi},\quad
    \varphi\in[0,2\pi).
\end{align}
As will soon become clear: this function is ideally suited to describing the visibility of the curve $\mathcal{C}$.
For now, we merely note that $f(\varphi)$ completely encodes the curve $\mathcal{C}$, which may be reconstructed via the inverse relations \cite{GrallaLupsasca2020c}
\begin{align}
    \alpha(\varphi)=f(\varphi)\cos{\varphi}-f'(\varphi)\sin{\varphi},\quad
    \beta(\varphi)=f(\varphi)\sin{\varphi}+f'(\varphi)\cos{\varphi}.
\end{align}
It will turn out to be useful to decompose the projected position into parity-even and parity-odd components,
\begin{align}
    \label{eq:DiameterDefinition}
    d_\varphi=f(\varphi)+f(\varphi+\pi),\quad
    C_\varphi=\frac{f(\varphi)-f(\varphi+\pi)}{2},\quad
    \varphi\in[0,\pi).
\end{align}
Geometrically, these components may be interpreted as the projected diameter $d_\varphi$ and projected centroid $C_\varphi$ of the curve when viewed from an angle $\varphi$ (see Fig.~3 of Gralla \& Lupsasca \cite{GrallaLupsasca2020c} for an illustration).

Gralla \cite{Gralla2020} showed that to leading order in its small width-to-diameter ratio $0<w/d\ll1$, a thin ring of generic (but convex) shape $\mathcal{C}$ produces in the universal regime \eqref{eq:UniversalRegime} a ``universal interferometric signature''\footnote{Recently, He et al \cite{Jia2024} derived the subleading corrections in $w/d$ to this expression.
They involve polynomials in $u$ and provide better fits for the visibility of the $n=1$ ring, for which $w/d\sim10\%$, but are not needed for the present discussion.}
\begin{align}
	\label{eq:UniversalVisibility}
	V(\bm{u})=\frac{e^{-2\pi iC_\varphi u}}{\sqrt{u}}\br{\alpha_\varphi^{\rm L}e^{-\frac{i\pi}{4}}e^{i\pi d_\varphi u}+\alpha_\varphi^{\rm R}e^{\frac{i\pi}{4}}e^{-i\pi d_\varphi u}},
\end{align}
where $\bm{u}=(u,\varphi)$ are polar baseline-plane coordinates, and $(d_\varphi,C_\varphi)$ are the projected diameter and centroid of the thin ring, while the functions $(\alpha_\varphi^{\rm L},\alpha_\varphi^{\rm R})$ encode its angular intensity profile.
Again, the radial profile of the ring (including the fact that it is not an infinitely thin curve) is irrelevant: all thin rings tracking the same shape $\mathcal{C}$ share the same visibility \eqref{eq:UniversalVisibility}, whence its label as a ``universal'' formula.

In practice, the phase of the complex visibility is more sensitive to measurement errors and therefore harder to measure than its amplitude.
In the universal regime \eqref{eq:UniversalRegime}, the amplitude of the universal ring visibility \eqref{eq:UniversalVisibility} takes the even simpler universal form
\begin{align}	
	\label{eq:UniversalVisamp}
	\ab{V(\bm{u})}=\frac{1}{\sqrt{u}}\sqrt{\pa{\alpha_\varphi^{\rm L}}^2+\pa{\alpha_\varphi^{\rm R}}^2+2\alpha_\varphi^{\rm L}\alpha_\varphi^{\rm R}\sin\pa{2\pi d_\varphi u}},
\end{align}
which depends only on the ring diameter $d_\varphi$ illustrated in the left panel of Fig.~\ref{fig:Diameter}, and no longer on its centroid.

Thus, by accessing the baselines \eqref{eq:UniversalRegime} dominated by the universal interferometric signature of a photon ring, one could in principle measure its angle-dependent diameter $d_\varphi$ at every angle, as illustrated in the middle panel of Fig.~\ref{fig:Diameter}.
This measurement could then be compared against the GR-predicted photon ring shape, to be described in detail in Sec.~\ref{subsec:PhotonRingShape} below.

Before doing so, we emphasize that a single space element is in principle sufficient to carry out a full measurement of $d_\varphi$.
As shown in the right panel of Fig.~\ref{fig:Diameter}, a satellite orbiting the Earth in the plane perpendicular to the line of sight can sample the visibility amplitude \eqref{eq:UniversalVisamp} at many angles $\varphi$ around its orbit.
In this simple cartoon picture, the periodicity of the interferometric ringing at orbital angles $\varphi$ and $\varphi+\pi$ is set by the diameter $d_\varphi$ of the ring at angle $\varphi$ around the image.

\subsection{Predicted interferometric signature of the photon ring}
\label{subsec:PhotonRingShape}

\begin{figure}[t!]
    \centering
    \includegraphics[width=\textwidth]{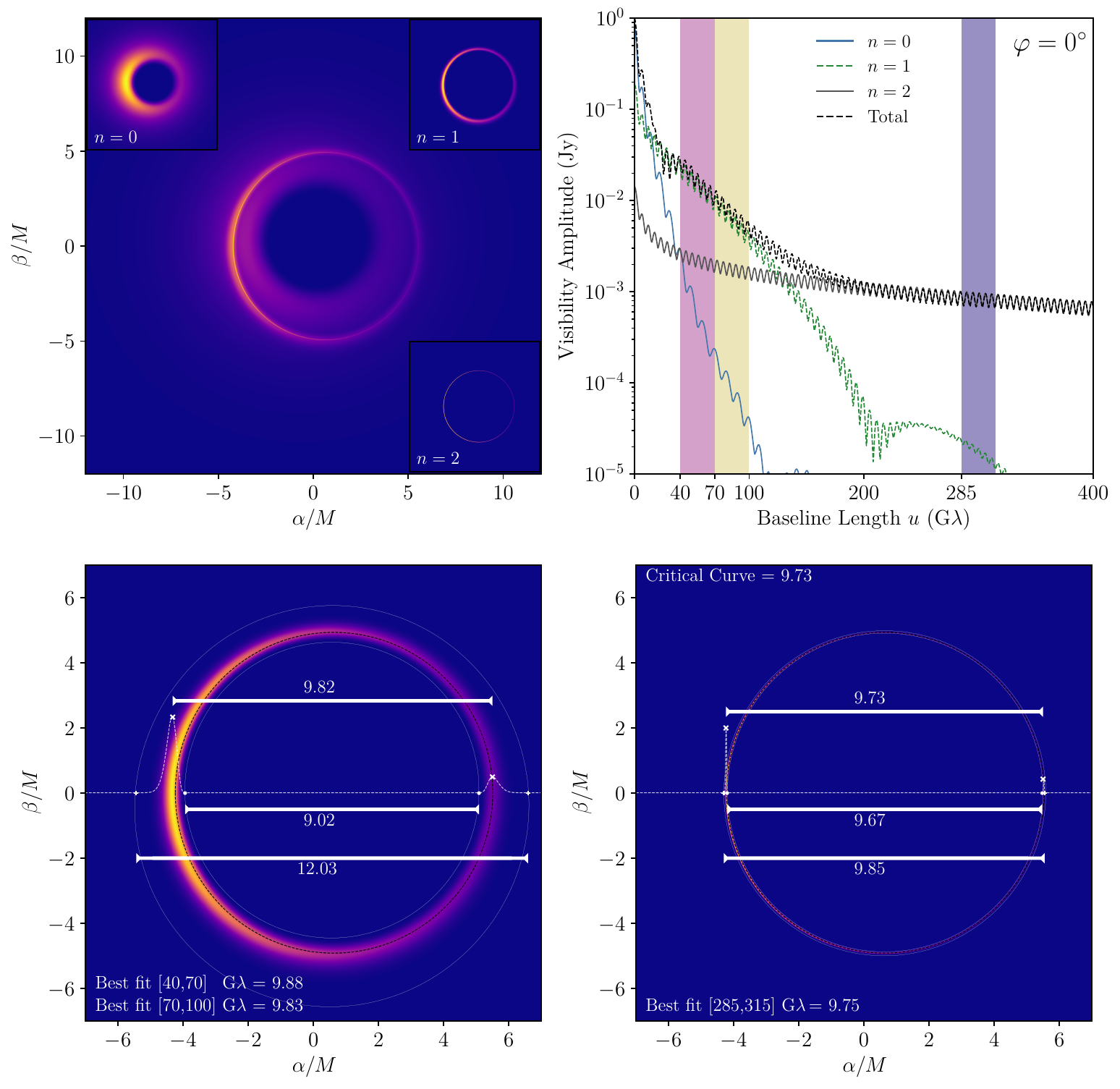}
    \caption{{\bf Time-averaged image and visibility of an equatorial source model for M87* decomposed into contributions from photons with definite half-orbit number $n$} (reproduced from Fig.~1 of C\'ardenas-Avenda\~no \& Lupsasca \cite{CardenasAvendano2023}).
    Top left: Image of a time-independent equatorial source model described in Sec.~\ref{sec:Model} ray traced with \texttt{AART} \cite{AART}.
    The inset panels show the decomposition of the full image into layers labeled by photon half-orbit number $n$.
    Top right: Corresponding visibility amplitude $\ab{V(u,\varphi)}$ at angle $\varphi=0^\circ$ in the baseline plane, decomposed into contributions from each image layer.
    Each subring dominates the full signal in its own domain \eqref{eq:DominantRing}, resulting in a universal interferometric ``cascade'' structure for the full photon ring.
    Bottom: The periodicity of the interferometric ringing in the regime dominated by the $n^\text{th}$ subring is controlled by its angle-dependent diameter $d_\varphi^{(n)}$ according to Eqs.~\eqref{eq:UniversalVisibility} and \eqref{eq:UniversalVisamp}.
    For extremely thin $n\gtrsim2$ subrings (right), this diameter is sharply defined in the image domain and the corresponding ringing signal subsists over many ``hops'' with the same periodicity.
    By contrast, for the $n=1$ subring (left), the diameter $d_\varphi^{(1)}$ is only defined in the image domain to within the width-to-diameter ratio $w/d\sim10\%$ of the ring.
    Correspondingly, the ringing visibility only describes a few hops of varying periodicity within the allowed range of image diameters.
    }
    \label{fig:Cascade}
\end{figure}

To complete the description of the interferometric signature of the photon ring, we must take into account its subring structure.
As discussed in Sec.~\ref{sec:RingImage}, the first ($n=1$) ring has characteristic width $w_1\sim M$, while each successive ring has an exponentially narrower width
\begin{align}
    w_{n+1}\sim e^{-n\gamma}w_1.
\end{align}
By the same argument as in Sec.~\ref{subsec:DeltaFunctionRing}, we expect the power contributed to the full visibility \eqref{eq:ComplexVisibility} by the $n^\text{th}$ subring to rapidly dwindle once the interferometer achieves the resolution $u\sim1/w_n$ needed to resolve its width.
Thereafter, its signal ought to be ``resolved out'' and become dominated by the next subring, whose exponentially narrower width has yet to be resolved.
As first pointed out by Johnson et al \cite{JohnsonLupsasca2020} (see Fig.~5 therein), this implies that the $n^\text{th}$ photon ring dominates the radio visibility in the range
\begin{align}
	\label{eq:DominantRing}
	\frac{1}{w_{n-1}}\ll u\ll\frac{1}{w_n}.
\end{align}
As a result, the full visibility consists of a ``cascade'' of damped oscillations with periodicity encoding the diameter $d_\varphi^{(n)}$ of successive subrings, as illustrated in Fig.~\ref{fig:Cascade}; for more discussion, see also Secs.~2 and 4 of Paugnat et al \cite{Paugnat2022}.

In summary, we expect the visibility amplitude of a black hole image to take the universal form \eqref{eq:UniversalVisamp} governed by the diameter $d_\varphi^{(n)}$ of the $n^\text{th}$ photon ring in the universal regime \eqref{eq:UniversalRegime} where it dominates the signal.
As discussed below Eq.~\eqref{eq:InfiniteSubring}, the photon rings converge (exponentially fast in $n$) to the critical curve $\tilde{\mathcal{C}}$, so their projected diameters $d_\varphi^{(n)}$ must also converge to that of the critical curve,
\begin{align}
    \tilde{d}_\varphi=d_\varphi^{(\infty)}
    \equiv\lim_{n\to\infty}d_\varphi^{(n)}.
\end{align}
The projected diameter $\tilde{d}_\varphi(M,a,\theta_{\rm o})$ of the critical curve was analyzed by Gralla \& Lupsasca \cite{GrallaLupsasca2020c}, who showed that it is extremely well-approximated by the ``circlipse'' shape
\begin{align}
	\label{eq:Circlipse}
	\frac{d_\varphi}{2}=R_0+\sqrt{R_1^2\sin^2\pa{\varphi-\varphi_0}+R_2^2\cos^2\pa{\varphi-\varphi_0}},
\end{align}
thus named because it is the sum of the diameters of a circle with radius $R_0$ and an ellipse with semi-major and semi-minor axes $R_1$ and $R_2$.
In practice, the overall angle $\varphi_0$ is needed to account for the rotation of the curve relative to the image-plane coordinate system, whose orientation is a priori unknown in actual observations \cite{GLM2020}.

As reviewed in Sec.~\ref{sec:RingImage}, the critical curve (or ``$n\to\infty$ subring'') is not in itself observable, so to make contact with experiment, one must instead obtain a GR prediction for the photon ring diameters $d_\varphi^{(n)}$ at finite $n$.
For large $n$, this diameter must track the circlipse shape \eqref{eq:Circlipse} of the critical curve, up to small corrections that are exponentially suppressed in $n$.
For $n=2$, however, precision measurements on very long baselines to lunar distances could in principle detect an observable deviation in $d_\varphi^{(2)}$ relative to $\tilde{d}_\varphi$ \cite{GLM2020}.
Nonetheless, Gralla et al \cite{GLM2020} found that the diameter of the $n=2$ ring must still follow the functional form \eqref{eq:Circlipse} to better than a part in one thousand, albeit with different parameters $(R_0,R_1,R_2,\varphi_0)$ than the critical curve diameter $\tilde{d}_\varphi$.
More surprisingly, C\'ardenas-Avenda\~no \& Lupsasca \cite{CardenasAvendano2023} later arrived at a similar conclusion for the first $n=1$ ring, despite its projected diameter $d_\varphi^{(1)}$ not even being sharply defined in the image domain.

In conclusion, we may regard the circlipse shape \eqref{eq:Circlipse} to be the GR-predicted functional form for the photon ring diameters $d_\varphi^{(n)}$, as illustrated in the middle panel of Fig.~\ref{fig:Diameter}.
The BHEX mission will target observations of M87* and Sgr\,A* in the baseline regime \eqref{eq:DominantRing} dominated by the $n=1$ ring, and extract the projected ring diameter $d_\varphi^{(1)}$ at many angles around its image (or equivalently, around the orbit).

Finally, we stress that the map between the black hole parameters (the mass $M$, spin $a$, and inclination $\theta_{\rm o}$) and the best-fit circlipse parameters $(R_0,R_1,R_2,\varphi_0)$ for the diameter $d_\varphi^{(n)}$ depends on the emission profile, with this dependence vanishing exponentially fast in $n$.
For the $n=1$ ring, the remaining astrophysical dependence places some limits on the precision of the parameter estimates that can be derived from purely geometric considerations alone, as discussed in Sec.~\ref{subsec:LensingBands}.

\subsection{Orbital constraints for photon ring measurements with BHEX}
\label{subsec:Constraints}

\begin{figure}[t]
    \centering
    \includegraphics[width=0.45\textwidth]{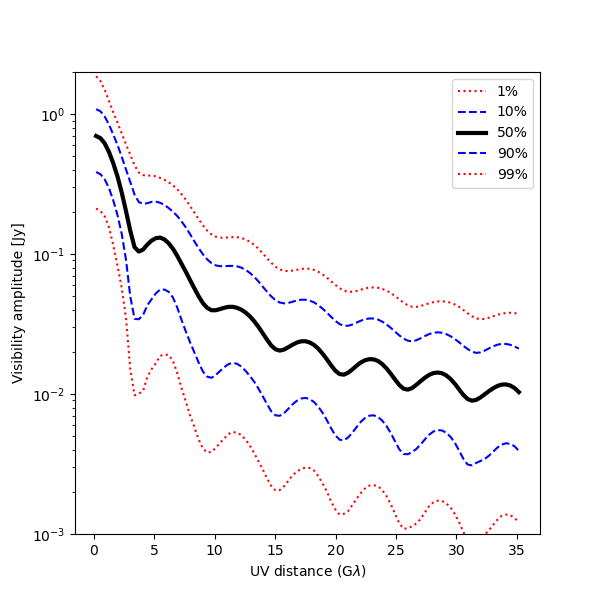}
    \includegraphics[width=0.45\textwidth]{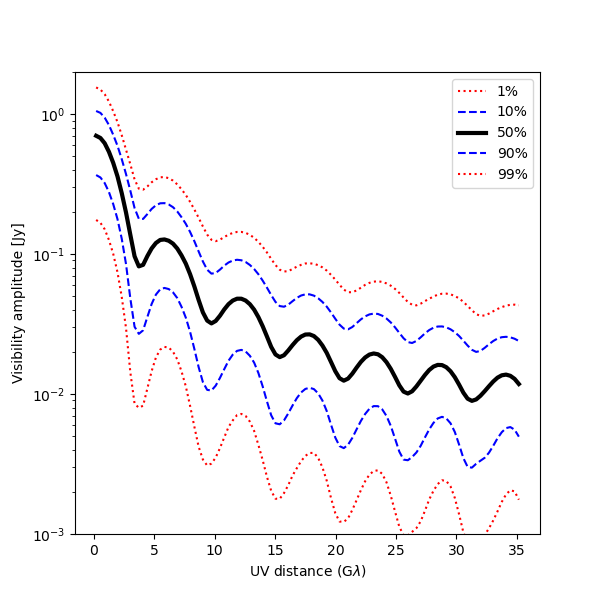}
    \caption{{\bf Quantiles of the visibility amplitude of M87* at 230\,GHz, averaged over many snapshots from multiple GRMHD simulations, and averaged at all angles around the baseline plane.}
    Plots computed by C.~Gammie and A.~Joshi from a large subset of GRMHD simulations in the Illinois library.
    Left: Average over both MAD and SANE models.
    Right: Average over only MAD states (favored by EHT observations of M87*).
    The flux density on BHEX baselines always exceeds 1\,mJy and oscillates around $|V(u)|\sim10\,$mJy in 90\% of models.
    }
    \label{fig:Quantiles}
\end{figure}

At this stage, we can already place some basic constraints on the orbital parameters of the BHEX satellite if it is to detect the $n=1$ photon rings around M87* and Sgr\,A*.
We note that the estimated ring diameters for M87* and Sgr\,A* are $d_{\rm M87^*}^{(1)}\approx42\,\mu$as and $d_{\rm Sgr\,A^*}^{(1)}\approx50\,\mu$as, respectively \cite{EHT2019a,EHT2022a}, corresponding to an interferometric ringing with visibility amplitude ``hops'' of $\Delta u_{\rm M87^*}\sim4.9\,$G$\lambda$ and $\Delta u_{\rm Sgr\,A^*}\sim4.1\,$G$\lambda$, respectively.

First, the BHEX mission must ensure that these rings are present in images taken at its observing frequency.
The EHT observations of M87* \cite{EHT2019a} and Sgr\,A* \cite{EHT2022a} strongly suggest that these sources are sufficiently optically thin for the $n=1$ ring to appear in 230\,GHz images.
This claim is also supported by GRMHD simulations, as shown in Fig.~\ref{fig:OpticalDepth}.
To completely retire the risk of the $n=1$ ring being obscured by optical depth, BHEX will push up its observing frequency as high as 300\,GHz \cite{Johnson2024,Marrone2024,Issaoun2024}.

Second, as shown in Fig.~\ref{fig:GravityStamp}, to convincingly detect a photon ring around M87*, the baselines targeted by BHEX must extend as far as $\sim20\,$G$\lambda$.
At the nominal EHT frequency of 230\,GHz, this requires an altitude of $\sim$26,000\,km above the ground in the plane perpendicular to the line of sight to M87*.
At the highest BHEX frequency of 300\,GHz, this is only an orbital distance of $\sim$20,000\,km.
Fortuitously, there exist favorable orbits with desirable characteristics for beaming data down to Earth \cite{Johnson2024,Marrone2024,Issaoun2024}, such as the one considered in Sec.~\ref{sec:GRMHD} below.

Third, the source must produce sufficient power in the visibility amplitude on long space baselines for the BHEX instruments to be capable of detecting its ringing signature.
Of course, the radio visibilities of M87* and Sgr\,A* have never been observed on such long baselines, so their flux densities there are unknown.
Nevertheless, we can verify that the flux density is sufficiently high in GRMHD models of M87*.
Averaging over all baseline angles at a given baseline radius, and then averaging over many snapshots from multiple GRMHD simulations (normalized such that $V(0)$ matches the compact flux of the source), one obtains the quantiles shown in Fig.~\ref{fig:Quantiles}.

\section{The photon ring in the presence of astrophysical fluctuations and instrument noise}
\label{sec:Model}

Thus far, the discussion in this paper has focused on static sources, or time-averaged images of time-dependent sources.
In GRMHD simulations, snapshot images display not only a photon ring (with its subring substructure) but also other transient features (such as plasma flares or emission ropes) which may ``mimic'' the photon ring.
If one averages snapshot images taken over a sufficiently long time, then these random fluctuations ought to wash out and leave behind only a pristine photon ring, which must always be there simply because it is the part of the image that belongs to the black hole itself---which is always there---rather than its circulating plasma.

However, this raises the question: how long is a ``sufficiently long'' time average?
Realistic observations such as those envisioned with BHEX must grapple with this.
Before tackling this question in the considerably more complicated context of GRMHD simulations, it is helpful to build some basic intuition using simple toy models.

To this end, here we briefly review a semi-analytic model of equatorial emission that can capture several of the key ingredients of black hole images without being computationally expensive.
This model was first introduced by Gralla et al \cite{GLM2020} to produce images whose intensity profile qualitatively matches the time-averaged intensity seen in GRMHD-simulated images such as the one shown in Fig.~\ref{fig:RadialProfile} (however, their focus was on futuristic observations of the $n=2$ ring using space-VLBI with a satellite orbiting at lunar distance, rather than the $n=1$ ring measurements currently envisioned with BHEX).
Soon after, Chael et al \cite{Chael2021} showed that with a suitable choice of radial emission profile in the equatorial plane, this match can even be made to agree quantitatively, leading to the more systematic investigation of the model by Paugnat et al \cite{Paugnat2022}.

Thereafter, this model was implemented in the relativistic ray tracer \texttt{AART} \cite{AART}, which uses an explicit analytical form of the null geodesic equation in the Kerr spacetime \cite{GrallaLupsasca2020b} to efficiently produce high-resolution images and their corresponding visibility.
The original version of the model assumed that the source consists of emitters traveling on circular-equatorial (Keplerian) geodesics.
Its implementation in \texttt{AART} generalizes it to include emitters on infalling trajectories, on (non-geodesic) sub-Keplerian circular-equatorial orbits, or on some combination thereof.
It also allows for the equatorial source to vary in time by calling upon the \texttt{inoisy} code \cite{Lee2021} to generate stochastic realizations of a Gaussian random field with correlation statistics prescribed by a M\'atern-like covariance.
For further details, we refer the reader to the \texttt{AART} code paper \cite{AART} and to the recent surveys based on it \cite{CardenasAvendano2023,CardenasAvendano2024}.

In a nutshell: in this model, the image intensity observed at time $t_{\rm o}$ in the image plane $(\alpha,\beta)$ of a distant observer at large radius $r_{\rm o}\gg M$, azimuthal angle $\phi_{\rm o}=0^\circ$, and polar inclination $\theta_{\rm o}$ is computed via the formula
\begin{align}
	\label{eq:IntensityProfile}
	I_{\rm o}(t_{\rm o},\alpha,\beta)=\sum_{n=0}^{N(\alpha,\beta)-1}\zeta_ng^3\pa{r_{\rm s}^{(n)},\alpha,\beta}I_{\rm s}\pa{r_{\rm s}^{(n)},\phi_{\rm s}^{(n)}, t_{\rm s}^{(n)}}.
\end{align}
Here, $g$ is the redshift factor of the observed emission, $\bm{x}_{\rm s}^{(n)}=\bm{x}_{\rm s}^{(n)}(\alpha,\beta)$ are the equatorial coordinates at which a ray shot backwards from the image plane in the direction $(\alpha,\beta)$ intersects the equatorial plane for the $(n+1)^\text{th}$ time along its trajectory, up to a total number $N(\alpha,\beta)$ along its maximal extension, and $I_{\rm s}$ denotes the intensity at the source, which can consist of either a radial emission profile or the output of an \texttt{inoisy} simulation.
Since this model neglects the geometrical thickness of a GRMHD accretion flow, to reproduce its time-averaged images, one must add a factor $\zeta_n$ that is assumed to be equal to $1$ for $n=0$, and $1.5$ for $n\geq1$ \cite{Chael2021,Vincent2022}.

\begin{figure}[t]
    \centering
    \includegraphics[width=\textwidth]{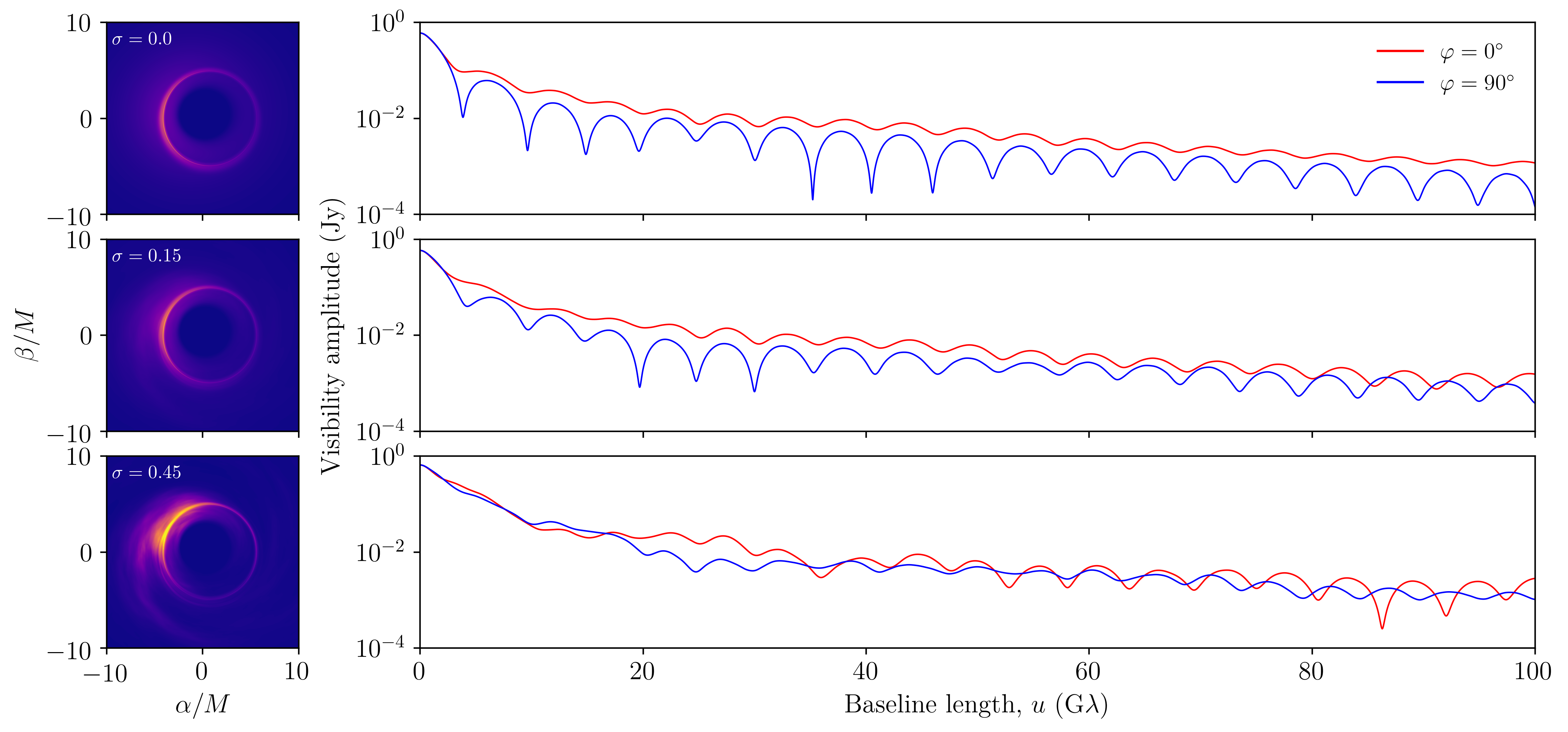}
    \vspace{4pt}\\
    \caption{{\bf Black hole snapshot images (left) and their respective visibility amplitudes (right) for spin-aligned (blue) and spin-perpendicular (red) cuts across each image} (reproduced from Fig.~1 of C\'ardenas-Avenda\~no et al \cite{CardenasAvendano2024}).
    From top to bottom, the characteristic scale $\sigma$ of astrophysical fluctuations in the source increases.
    In the absence of such fluctuations ($\sigma=0$), each snapshot image is the same as the time-averaged image, with an underlying radial emission profile of the form \eqref{eq:IntensityProfile}.
    The visibility amplitudes clearly show that as the amplitude ($\sigma=0$) of the astrophysical fluctuations grows, it becomes strictly harder to identify a clear ringing in the interferometric signal.
    Nevertheless, by averaging repeated observations with sufficient sensitivity, it is possible to recover a periodic signal from which the $n=1$ ring diameter can be extracted.
    }
    \label{fig:Snapshots}
\end{figure}

Without going into the details, we briefly summarize the main takeaways from the most recent \texttt{AART} survey \cite{CardenasAvendano2024}:
\begin{itemize}
    \item The presence of astrophysical fluctuations with a characteristic scale $\sigma$ tends to attenuate the periodic interferometric ringing signature of the photon ring as $\sigma$ grows (Fig.~\ref{fig:Snapshots}).
    However, by averaging over $N$ snapshot images, the clear ringing pattern and its periodicity can be quickly recovered, with already precise measurements for moderately small $N\sim 5$.
    Surprisingly, this is true for both coherent averages (in which one first averages the complex visibility \eqref{eq:ComplexVisibility} of all snapshots before taking the amplitude of the average) {\it and} for incoherent averages (in which one does the reverse, first taking the amplitude of each snapshot visibility and then averaging them).
    Such averages do not commute, and the visibility of a time-averaged image is a coherent average, which is much harder to perform in practice.
    The incoherent average is more easily attained but could in principle fail to display a clear interferometric signature associated with the photon ring, even if that is not the case across the examples considered across the survey \cite{CardenasAvendano2024}.
    \item The photon ring diameter can be measured even in the presence of instrument noise across a wide range of baselines.
    The measurement accuracy and precision appear relatively insensitive to the noise level (modeled as complex Gaussian noise \cite{CardenasAvendano2024}), up to a sharp threshold beyond which any measurement becomes incredibly challenging (at least without recourse to more sophisticated data analysis methods).
\end{itemize}
With these encouraging results in hand, we finally turn to the forecasting of BHEX observations over one night.

\section{The photon ring of M87* as observed over one night with BHEX}
\label{sec:GRMHD}

Before a detailed discussion of data analysis techniques and feature extraction algorithms, a few simple constraints (already mentioned in Sec.~\ref{subsec:Constraints}) help to determine the range of feasible mission architectures that efficiently enable detection of the photon rings in M87* and Sgr\,A*.
First, each source must be observed at a frequency such that the accretion flow is optically thin on photon trajectories corresponding to the $n=1$ ring (Fig.~\ref{fig:OpticalDepth}).
Second, each source must be observed on baselines that are sufficiently long to be dominated by the interferometric signature of the $n=1$ ring (Fig.~\ref{fig:GravityStamp}).
Third, BHEX must be sensitive enough to detect the interferometric visibility signal from the source (Fig.~\ref{fig:Quantiles}).
Since interferometric visibilities decay as baseline lengths increase, these considerations impose a joint constraint on the instrument sensitivity and typical baseline lengths from Earth sites to the source.

Additional morphological considerations also constrain the possible choices of orbit.
The shape of the photon ring encodes rich information about the black hole spin and viewing inclination; this shape information is only accessible if the interferometric baselines to BHEX sweep through a wide range of angles in the Fourier domain, capturing many projections of the source structure \cite{Bracewell1956,GLM2020}.
This constraint is particularly important for M87*, where the long dynamical time ($\sim$days) permits thorough sampling of the Fourier plane in a static-image limit.

For Sgr\,A*, the morphological considerations are dominated by scattering off the interstellar plasma, which manifests as both a diffractive ``blurring'' effect that reduces the signal on long baselines, and a refractive ``noise'' effect that adds fine structure to the image \cite{Rickett1990,Narayan1992,JohnsonNarayan2016,Johnson2016}.
The impact of scattering increases with the square of the observed wavelength; in order to observe the photon ring around Sgr\,A*, BHEX must observe at a frequency such that the refractive scattering does not outshine the photon ring on Earth-BHEX baselines.
In addition, scattering is anisotropic, with a significantly greater impact on long east-west baselines than north-south.
Recent simulations \cite{Palumbo2023} have found that, despite observations of Sgr\,A* at 230\,GHz never being photon-ring-dominated due to refractive scattering, north-south interferometric observations at $\sim300$\,GHz show promise for revealing the Sgr\,A* photon ring.

\begin{figure}[t]
    \includegraphics[width=\textwidth]{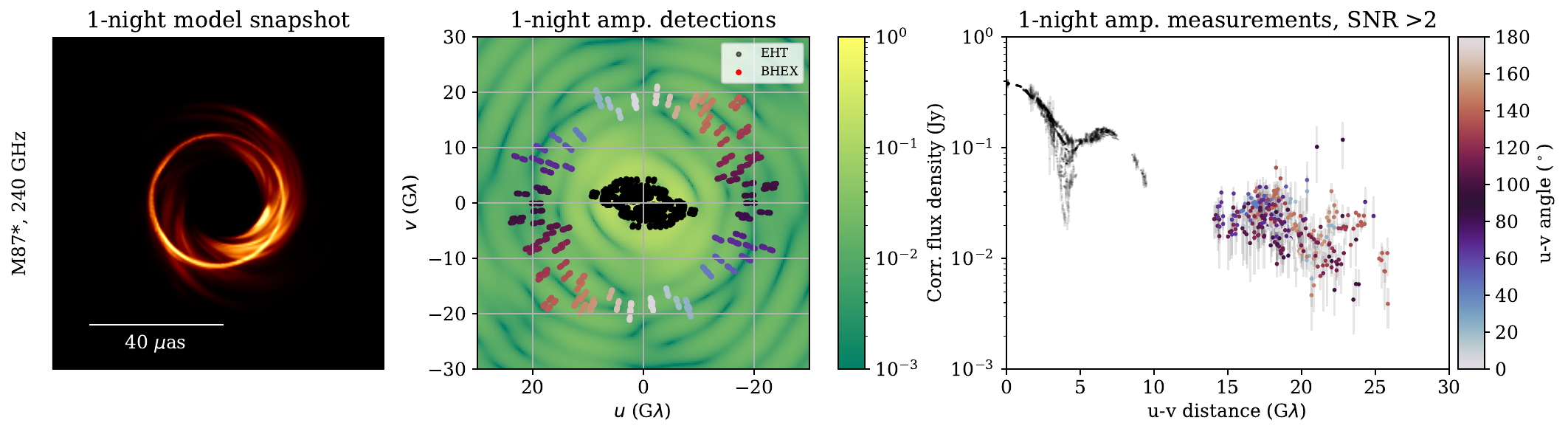}
    \caption{{\bf Simulation of a night of M87* observations using BHEX, including effects from accretion physics to Earth weather.}
    Forecasts predict detections on visibility amplitudes across the Fourier plane throughout winter.
    Left: Snapshot image of M87* GRMHD accretion flow at 240\,GHz.
    Middle: Corresponding Fourier amplitudes, with detections in both BHEX sidebands overplotted.
    Right: Measured amplitudes with telltale oscillations of the photon ring.}%The assumed ground array includes the IRAM 30m, JCMT, GLT, Haystack, Kitt Peak, LMT, NOEMA, and SMA.
    \label{fig:BHEX}
\end{figure}

Taken together, these constraints favor orbits that appear circular when projected to M87* and elliptical along the north-south axis when projected to Sgr\,A*.
The size of the orbit is set primarily by the expected source signal, but also by the eventual domination of refractive scattering in Sgr\,A* on baselines that are too long, even at 345\,GHz.
GRMHD simulations provide the best available estimate of source brightness and morphology, and suggest that the photon ring will dominate the signal while still being detectable in the range of $\sim20\,$G$\lambda$ to $40\,$G$\lambda$ for both sources.
Astrophysical uncertainty about the details of the accretion flow and emission geometry strongly favor frequency flexibility; there are reasonable models for M87* for which detection of the photon ring is a major challenge above 300\,GHz, but are safely detectable closer to 200\,GHz.
Meanwhile, the photon ring around Sgr\,A* may be impossible to detect below 300\,GHz. 

Thus, the baseline instrument specifications for BHEX involve a polar orbit with a 12-hour period observing simultaneously in a low band (80 \,GHz to 100\,GHz) and a high band (tunable from approximately 240\,GHz to 320\,GHz) \cite{Marrone2024}.
In Fig.~\ref{fig:BHEX}, we show an example of a night of observations of M87* in the lower 8\,GHz BHEX sideband using weather and instrument modeling from \texttt{ngEHTsim} \cite{Pesce2024}.
The GRMHD model image is from a moderately spinning ($a_*=+0.5$) magnetically arrested disk simulated with the code \texttt{HARM3d} \cite{Gammie2003,Prather2021} with the electron heating parameter $R_{\rm high}=80$ \cite{Moscibrodzka2016}, ray traced using the \texttt{IPOLE} code \cite{Moscibrodzka2018,Wong2022}.
This pipeline constitutes an end-to-end treatment of the GR and astrophysical effects at the source as well as the observational details on the ground, and predicts BHEX observations rich with information about the black hole photon ring.

\section{Discussion}
\label{sec:Discussion}

We have given a review of the photon ring science that can be carried out with BHEX.
We paid special attention to the interferometric signature of the photon ring, and in particular how the visibility amplitude (which is most amenable to observation) on long baselines encodes the projected diameter of the photon ring.
We have argued that this GR-predicted angle-dependent diameter can be extracted from sparse samples of the visibility as will be measured by BHEX.
This paves the way to future tests of the Kerr hypothesis and precision measurements of black hole mass and spin using space-VLBI observations from an orbiting BHEX satellite to a ground array.

In 1973, Bardeen wrote (on p231 of his Les Houches lecture notes \cite{Bardeen1973}) in reference to his pioneering derivation of the critical curve that
\begin{quote}
    ``It is conceptually interesting, if not astrophysically very important, to calculate the precise apparent shape of the black hole.'' 
\end{quote}
One year later, he wrote (on p139 of a paper \cite{Bardeen1974}) that
\begin{quote}
    ``Unfortunately, there seems to be no hope of observing this effect.''
\end{quote}
Exactly half a century later, the EHT has released horizon-scale images of not one but two supermassive black holes, and BHEX is now proposing precision measurements of the photon ring from space.
Black hole imaging has a bright future.

\acknowledgments 

We are grateful to Andrew Chael, Charles Gammie, Joseph Farah, Shahar Hadar, Abhishek Joshi, Daniel Kapec, Eliot Quataert, Andrew Strominger, Maciek Wielgus, and George Wong for useful discussions about this work.

This work was supported in part by the National Science Foundation (grants AST-2307887, AST-2307888 and PHY-2340457) and by the Black Hole Initiative at Harvard University, which is funded by grant 62286 from the John Templeton Foundation and grant 8273.01 from the Gordon and Betty Moore Foundation.
In addition, {\bf BHEX is supported by initial funding from Fred Ehrsam.}

\appendix

\section{Frequently asked questions}

In this appendix, we clarify some common confusions and collect answers to a few frequently asked questions.

\subsection{Will BHEX measure Hawking radiation?}

{\bf Answer:} No.
The light observed in black hole images is mostly synchrotron radiation emitted from hot gas accreting onto the black hole.
This emission has a characteristic temperature of billions of degrees Kelvin (Fig.~\ref{fig:RadialProfile}):
\begin{align}
    T_{\rm gas}\approx\times10^{10}\,{\rm K}.
\end{align}
By contrast, the Hawking formula for the blackbody temperature of a non-rotating (Schwarzschild) black hole is
\begin{align}
    T_{\rm H}=\frac{\hbar c^3}{8\pi k_{\rm B}G_{\rm N}M},
\end{align}
which scales like the inverse mass.
For a supermassive black hole like M87* with mass $M\approx6.4\times10^8M_\odot$, this is
\begin{align}
    T_{\rm H}\approx6\times10^{-17}\,{\rm K}.
\end{align}
Thus, the black hole is akin to a very cool object in a hot room, and its radiation cannot escape the plasma.
With such a minute temperature, the blackbody radiation of the black hole would be unobservable with modern technology, even from the near-vicinity of the black hole.

\subsection{The photon ring is ``quantized'' into subrings---is this a quantum effect?}

{\bf Answer:} No.
The photon ring does have a (universal) discrete substructure, described in Sec.~\ref{subsec:Criticality} above.
As discussed therein, this structure is caused by the presence of ``gaps'' in the emission region near the north and south poles of the black hole: the bulk of the radiation observed in simulated images (such as the one in Fig.~\ref{fig:RadialProfile}) is emitted by near-equatorial sources, whose observational appearance consists of a discrete sequence of strongly lensed images.
This ``quantization'' of the photon ring into discrete subrings is therefore a consequence of the geometry of the emission region, combined with the lensing behavior of the black hole.
It has nothing to do with the quantum properties of the black hole, which are roughly suppressed in the image by
\begin{align}
    \frac{T_{\rm H}}{T_{\rm gas}}\approx10^{26}.
\end{align}

\subsection{What is the black hole shadow and what does it have to do with the photon ring?}

{\bf Answer:} The ``black hole shadow'' is a term with two incompatible but commonly used meanings:
\begin{itemize}
    \item On the one hand, it can refer to the ``critical curve'' $\tilde{\mathcal{C}}$ of a Kerr black hole, whose analytical form is explicitly given in Eq.~\eqref{eq:CriticalCurve}.
    As reviewed in Sec.~\ref{subsec:Criticality}, the critical curve is completely determined by the black hole spacetime geometry, and therefore encodes its parameters: the black hole mass, spin, and inclination.
    Unfortunately, it is a purely mathematical curve and is not in itself directly observable, except in some special scenarios in which the source is fine-tuned to produce a brightness deficit in its image that happens to coincide exactly with the interior of $\tilde{\mathcal{C}}$ \cite{Falcke2000,Gralla2019,Narayan2019}.
    \item On the other hand, it can also refer to a characteristic feature of black hole images: the central brightness depression caused by the presence of a black hole embedded within the source (Fig.~\ref{fig:PhotonRingStack}).
    This definition is generically not consistent with the previous one because the central brightness deficit does not typically coincide with the exact interior of the critical curve $\tilde{\mathcal{C}}$.
\end{itemize}
Historically, the spherical accretion scenarios in which the brightness deficit in the full image coincides with the critical curve were investigated first, so these two notions of ``shadow'' (namely, the interior of $\tilde{\mathcal{C}}$ and the central brightness depression observed in images) were sometimes conflated.
By contrast, the currently favored GRMHD models of M87* generically predict a near-equatorial source, for which these two notions are distinct:
\begin{itemize}
    \item In such scenarios, the brightness deficit in the $n=0$ layer (which corresponds to the main emission) is the direct image of the equatorial event horizon.
    This region is completely distinct from the critical curve, and for this reason this image feature has been given a different name: the ``inner shadow'' of the black hole \cite{Chael2021}.
    The brightness deficit in the full image is the ``inner shadow'' and {\it not} the usual ``shadow'' image feature defined as a brightness deficit that fills the interior of the critical curve $\tilde{\mathcal{C}}$ (see left panel of Fig.~\ref{fig:PhotonRingStack}).
    \item At the same time, the brightness depression in the $n\geq1$ layers (corresponding to the strongly lensed images of the main emission) does approximately match the interior of the critical curve $\tilde{\mathcal{C}}$.
    This deficit is precisely the interior of the $n^\text{th}$ subring $\mathcal{C}_n$, which rapidly tends to the critical curve $\tilde{\mathcal{C}}=\mathcal{C}_\infty$.
\end{itemize}
Observations with BHEX on long baselines to space will extract the diameter of the $n=1$ ring, and of the ``shadow'' image feature that corresponds to the brightness deficit in its interior.
This diameter is close to, but does not exactly coincide with, that of the critical curve.

\bibliography{PhotonRingSPIE} 
\bibliographystyle{spiebib} 

\end{document}